\magnification = \magstep1
%\nopagenumbers
\hsize=15truecm
\hoffset=0.8truecm
\vsize=21.6truecm
\voffset=0.3truecm

\def\t#1{#1}
\def\t#1{\empty}

\def\dblbaselines{\baselineskip=15pt 
                     \lineskip=0pt
                     \lineskiplimit=0pt}
\def\vsl{\vskip\baselineskip}
\def\vs{\vskip 10pt}
\def\skipit{\hskip 7pt}
 % used in equation (8)
\parskip = 0pt % default is \parskip = 0pt+1pt
\def\ts{\thinspace}
\def\cl{\centerline}
\def\ni{\noindent}
\def\nhi{\noindent \hangindent=1.0truecm}
\def\nhhi{\noindent \hangindent=3.30truecm}

\def\makeheadline{\vbox to 0pt{\vskip-30pt\line{\vbox to8.5pt{}\the
\headline}\vss}\nointerlineskip}

\def\footnoterule{\kern-3pt \hrule width \hsize \kern 2.6pt \vskip 3pt}
\output={\plainoutput}
\pretolerance=3000
\tolerance=5000
\def\sup1{$^{\rm 1}$}
\def\sup2{$^{\rm 2}$}
\def\r0{$\rho_0$}

\def\00{$\phantom{000000}$}
\def\0{\phantom{0}}
\def\1{\phantom{1}}

\def\c{{\it c}}\def\e{{\it e}}

\newdimen\sa \newdimen\sb
\def\sd{\sa=.1em 
           \ifmmode $\rlap{.}$''$\kern -\sa$
           \else \rlap{.}$''$\kern -\sa\fi}
\def\ss{\ifmmode ^{\prime\prime}$\kern-\sa$ \else $^{\prime\prime}$\kern-\sa\fi}
\def\mm{\ifmmode ^{\prime}$\kern-\sa$ \else $^{\prime}$\kern-\sa \fi}

\def\m31{M{\ts}31}
 
\cl{\null} \vsl
 
\cl{ SOURCES OF RELATIVISTIC JETS IN THE GALAXY}
\vskip 2.0truecm
 
\cl{\it I. F. Mirabel}
\vs
\cl {Centre d'Etudes de Saclay, CEA/DSM/DAPNIA/Sap}
\vs
\cl {F-91191 Gif-sur-Yvette, France, and}
\vs 
\cl {Instituto de Astronom\'\i a y F\'\i sica del Espacio}
\vs
\cl {C.C. 67, Suc. 28. 1428, Buenos Aires, Argentina}
\vs
\cl {mirabel@discovery.saclay.cea.fr}
 
\vsl
\cl {and}
\vsl
 
\cl{\it L. F. Rodr\'\i guez}
\vs
\cl {Instituto de Astronom{\'\i}a, UNAM,}
\vs
\cl {Apdo. Postal 70-264,
              04510 M\'exico, D.F., M\'exico}
\vs
\cl {luisfr@astrosmo.unam.mx}
 
\vfill

\ni  {\it Shortened title:} \skipit RELATIVISTIC JETS IN THE GALAXY
\vsl
\nhhi {\it Send Proofs to: I. F. Mirabel (please contact by email
first to check for address)}
 
\vsl
\ni {\it KEY WORDS: radio continuum stars, superluminal motion, X-rays binaries}

\vsl\vsl
 
\dblbaselines
 
%\hrule
%\vskip 4pt
%$^1${\ts}Also at Instituto de Astronom\'\i a
%y F\'\i sica del Espacio. C.C. 67, Suc. 28. 1428, 
%Buenos Aires, Argentina.

\eject
%\end
\ni ABSTRACT 
\vsl

Black holes of stellar mass and neutron stars 
in binary systems are first detected as
hard X-ray sources using high-energy space telescopes.
Relativistic jets in some of these compact
sources are found by means of  
multiwavelength observations with ground-based telescopes.
The X-ray emission 
probes the inner accretion disk and immediate surroundings of the compact object, 
whereas the synchrotron emission from the jets
is observed in the radio and infrared bands, and in the future could be detected
at even shorter wavelengths.
Black-hole X-ray binaries with relativistic
jets mimic, on a much smaller scale,
many of the phenomena seen in quasars and are
thus called microquasars.
Because of their proximity, their study opens the way for a better 
understanding of the relativistic jets seen elsewhere in the Universe. 
From the observation of two-sided moving jets it is inferred that 
the ejecta in microquasars
move with relativistic speeds similar to those believed to be present in quasars. 
The simultaneous multiwavelength approach to microquasars reveals in short 
timescales the 
close connection between instabilities in the accretion disk seen in the 
X-rays, and the ejection of relativistic clouds of plasma observed as 
synchrotron emission at longer wavelengths. 
Besides contributing to a deeper comprehension 
of accretion disks and jets, microquasars may serve in the future to determine 
the distances of jet sources using constraints from special relativity, and the 
spin of black holes using general relativity.

\eject

%\pageno=3\toppageno
\cl{\null}
\vskip 0pt
 
\ni 1. JETS IN ASTROPHYSICS
\vsl
 
While the first evidence of jet-like features emanating from the nuclei of galaxies
goes back to the discovery by Curtis (1918) of the optical jet from 
the elliptical galaxy M87 in the Virgo
cluster, the finding that jets can also be produced in
smaller scale by binary stellar systems is much more recent. 
The detection by Margon et al. (1979)
of large, periodic Doppler drifts in the optical lines
of SS433 resulted in the proposition of a kinematic
model (Fabian \& Rees 1979; Milgrom 1979) consisting of two precessing jets of
collimated matter with velocity of 0.26c.
High angular radio imaging as a function of time showed the presence of
outflowing radio jets and fully confirmed the kinematic model
(Spencer 1979; Gilmore \& Seaquist 1980;
Gilmore et al. 1981; Hjellming \& Johnston 1981).
The early history of SS433 has been reviewed by Margon (1984).

Since the detection of Sco X-1 at radio wavelengths (Ables 1969),
some X-ray binaries had been known to be strong, time-variable non-thermal 
emitters.
Ejection of synchrotron-emitting clouds was suspected from 
those days, but the actual confirmation of radio jets came only with
the observations of SS433.
At present, there are about 200 known galactic X-ray binaries
(van Paradijs 1995), of which about 10 percent are radio-loud
(Hjellming \& Han 1995). Of these radio-emitting X ray binaries,
9 have shown evidence of relativistic jets
of synchrotron emission, and this
review focuses on this set of objects.
After the definition of
Bridle \& Perley (1984) for extragalactic jets, we 
use the term ``jets'' to designate
collimated ejecta
that have opening angles $\leq$15$^\circ$. 

In the last years it has become clear that collimated ejecta 
can be produced in several stellar environments
when an accretion disk is present. Jets
with terminal velocities in the order of a few hundred to a few thousand
km s$^{-1}$ are now known to emanate from objects as diverse as
very young stars (Reipurth \& Bertout 1997),  
nuclei of planetary nebulae (L\'opez 1997), and accreting white dwarfs 
that appear as supersoft X-ray sources (Motch 1998, Cowley et al. 1998).
These types of stellar jets have, however, non-relativistic
velocities ($\sim$100-10000 km s$^{-1})$
and their associated emission is dominantly
thermal (i.e. free-free continuum emission
in the radio
as well as characteristic near-IR, optical and UV lines).
Interestingly, in all known types of jet sources a disk is believed
to be present.
This review concentrates on synchrotron jets with
velocities that can be considered relativistic ($v~\geq~0.1c$),
which are observed in X-ray binaries
that contain a compact object, that is, a neutron star or a black hole.
Our emphasis is on the radio characteristics of these sources.
For detailed reviews of the X-ray properties of these sources
we refer the reader to the reviews by Tanaka \& Shibazaki (1996) and Zhang et al.
(1997).  

\vsl
\ni 2.~MICROQUASARS
\vsl

At first glance it may seem paradoxical that relativistic jets were 
first discovered in the nuclei of galaxies and distant quasars and that 
for more than a decade SS433 was the only known object of its class in 
our Galaxy (Margon 1984). The reason for this is that disks 
around supermassive black holes emit strongly at optical and UV wavelengths.
Indeed, the more massive
the black hole, the cooler the surrounding accretion disk is.
For a black hole accreting at the Eddington limit,
the characteristic black body  
temperature at the last stable orbit in the
surrounding accretion disk will
be given approximately by $T \sim 2 \times 10^7~M^{-1/4}$
(Rees 1984), with
$T$ in K and the mass of the black hole, $M$, in solar masses.
Then, while accretion disks in AGNs have strong emission in the 
optical and ultraviolet with distinct 
broad emission lines, black hole and neutron star 
binaries usually are identified for the first time by their X-ray emission. 
Among these sources, SS 433 is unusual given its broad optical emission lines
and its brightness in the visible. 
Therefore, it is understandable that 
there was an impasse in the discovery of new stellar sources of relativistic 
jets until the recent developments in X-ray astronomy. 
 
Observations in the two extremes of the electromagnetic 
spectrum, in the domain of the hard X-rays on one hand 
(Sunyaev et al. 1991; Paul et al. 1991), 
and in the domain 
of radio wavelengths on the other hand, revealed the existence of new 
stellar sources of relativistic jets known as {\it microquasars} 
(Mirabel et al. 1992; Mirabel \& Rodr\'\i guez 1998). These are stellar-mass black holes
in our Galaxy that mimic, on a smaller scale, many of the
phenomena seen in quasars.
The microquasars combine 
two relevant aspects of relativistic astrophysics: accreting black holes 
(of stellar origin) which are a prediction of general relativity and are 
identified by the production of hard X-rays 
and gamma-rays from surrounding accretion disks, and relativistic jets 
of particles that are understood in terms of special relativity and 
are observed by means of their synchrotron emission. 
 
Multi-wavelength studies of the X-ray and gamma-ray sources in 
the galactic center region led in the year 1992 to the discovery of two 
microquasars: 1E1740.7-2942 and GRS 1758-258 (Mirabel et al. 1992;
Rodr\'\i guez, Mirabel, \& Mart\'\i\ 1992). The X-ray luminosity, 
the photon spectrum, and the time variability of these two sources are comparable 
to those of the black hole binary Cygnus X-1
(Churazov et al. 1994; Kuznetsov et al. 1997), and it is unlikely that they are 
extragalactic since no such persistent hard X-ray ultraluminous AGNs 
are observed (Mirabel et al. 1993). 
In Figure 1 we show the radio counterpart of 1E1740.7-2942. As in 
Cygnus X-1, the centimeter
radio counterpart of 1E1740.7-2942 is a weak 
core source that exhibits flux variations of the order of $\sim$ 50\% which 
at epochs appear anticorrelated with the X-ray flux (Mirabel et al. 1992). 
At radio wavelengths these two X-ray persistent 
sources located near the galactic center have a striking morphological
resemblance with distant radio galaxies; they consist 
of  compact components at the center of two-sided jets that end in weak, extended  
lobes with no significant radio flux variations observed in the last 6 years 
(Rodr\'\i guez \& Mirabel 1999b). 
1E1740.7-2942 and GRS 1758-258 seem to be persistent sources of both X-rays 
and relativistic jets. Mirabel et al. (1993)
have argued why it would be unlikely that the radio sources are radio galaxies
accidentally superposed on the X-rays sources.
For 1E1740.7-2942 no counterpart in the optical or near 
infrared wavelengths has been found so far, although there is a report of
a marginal detection at $\lambda$3.8 $\mu$m 
% it may have been marginally
% detected at L'
by Djorgovski et al. (1992). GRS 1758-258 has two possible faint candidate 
counterparts (Mart\'\i\ et al. 1998). 
 
In these binaries of stellar-mass are found the three basic ingredients 
of quasars; a black hole, an accretion disk heated by viscous dissipation, 
and collimated jets of high energy particles. But in microquasars the black hole 
is only a few solar masses instead of several millon solar masses; the 
accretion disk has mean thermal temperatures of several millon degrees 
instead of several thousand degrees; and the particles ejected at relativistic 
speeds can travel up to distances of a few light years only, instead of 
several millon light years as in giant radio galaxies 
(Mirabel \& Rodr\'\i guez (1998). Indeed, simple
scaling laws govern the physics of flows around
black holes, with length and time scales being proportional
to the mass of the black holes (Sams et al. 1996; Rees 1998). The word {\it microquasar} 
was chosen to suggest that the analogy with quasars is more than morphological, and 
that there is an underlying unity in the physics of accreting black holes 
over an enormous range of scales, from stellar-mass black holes in binary 
systems, to supermassive black holes at the center of distant galaxies. 
Strictly speaking and not being for the historical circumstances 
described above, the acronym  {\it quasar} (``quasi-stellar-radio-source") 
would have suited better the stellar mass versions rather than their 
super-massive analogs at the centers of galaxies. 
 
\vsl
\ni 3.~SUPERLUMINAL SOURCES
\vsl

Expansions at up to ten or more times the speed of light 
have been observed in quasars for more than 20 years
(Pearson \& Zensus 1987; Zensus 1997). At first these superluminal motions 
provoked concern because they appeared to violate relativity, but 
they were soon interpreted as illusions due to relativistic aberration (Rees, 1966). 
However, the ultimate physical interpretation had remained uncertain. In the 
extragalactic case the moving jets are observed as one-sided 
(because strong Doppler favoritism is required to render the 
approaching ejecta detectable) and it is not 
possible to know if superluminal motions represent the propagation of waves 
through a slowly moving jet, or if they reflect the actual bulk motion of the 
sources of radiation. 
 
In the context of the microquasar 
analogy, one may ask if superluminal motions could be observed from sources 
known to be in our own Galaxy. 
Among the handful of black holes of stellar mass known so far, three 
transient X-ray sources  
have indeed been identified at radio waves as sporadic sources of 
superluminal jets. The first superluminal source to be discovered
(Mirabel \& Rodr\'\i guez 1994) was 
GRS 1915+105, a recurrent transient source of hard X-rays first found and studied 
with the satellite GRANAT (Castro-Tirado et al. 1994; Finogenov et al. 1994). 
The discovery of superluminal 
motions in GRS 1915+105 stimulated a search for similar relativistic 
ejecta in other transient hard X-ray sources. Soon after, 
%superluminal 
%motions were reported in GRS 1915+105 by Mirabel \& Rodr\'\i guez (1994), 
the same phenomenon was observed by two different groups (Tingay et al. 1995;
Hjellming \& Rupen 1995)  in 
GRO J1655-40, a hard X-ray nova found with the Compton Gamma Ray Observatory
(Zhang et al. 1994). A third superluminal source may be XTE J1748-288 
(Hjellming et al. 1998), a transient source with a hard X-ray spectrum recently found 
with XTE (Smith et al. 1998).
 
GRS 1915+105 is at $\sim$12 kpc from the Sun (Rodr\'\i guez et al. 1995; 
Chaty et al. 1996) on the opposite side of 
the galactic plane and cannot be studied in the optical.
Given the large extinction by dust along the line of sight (Mirabel et al. 1994; 
Chaty et al. 1996), the precise nature 
of the binary has been elusive. Castro-Tirado et al. (1996)
proposed that GRS 1915+105 
is a low mass binary, while
Mirabel et al. (1997) proposed that it 
is a long period binary with a companion star of 
transitional spectral type. From the nature of the line variability in 
the infrared, Eikenberry et al. (1998b) propose that the emission lines in 
GRS 1915+105 arise in an accretion disk rather than in the circumstellar disk 
of an Oe/Be companion (Mirabel et al. 1997). GRS 1915+105 has similarities 
in the X-rays and gamma-rays with GRO J1655-40 and other black hole 
binaries, and it is also likely to harbor a black hole (Greiner, Morgan,
\& Remillard 1996). The X-ray 
luminosity of GRS 1915+105 (reaching 2 10$^6$ solar luminosities) far 
exceeds the 
Eddington limit (above which the radiation pressure 
will catastrophically blow out the external layers of 
the source) for a 3 solar mass 
object, which is 10$^5$ solar luminosities. 
Furthermore, it shows the typical hard X-ray tail
beyond 100 keV seen in black hole binaries (Cordier 1993; 
Finogenov et al. 1994; Grove et al. 1998). Finally, it is 
known that the absolute hard X-ray luminosities in black hole
systems are systematically higher than in neutron star systems (Ballet et al. 1993; 
Barret, McClintock, \& Grindlay 1996), another
result that points to a black hole in GRS 1915+105.

GRO J1655-40 is at a distance of 3.2 kpc and
the apparent transverse motions of its ejecta
in the sky are the largest yet 
observed (Tingay et al 1995; Hjellming \& Rupen 1995) until 
now from an object beyond the solar system.
It has a bright optical counterpart and consists of a star of
1.7-3.3 solar masses orbiting around a collapsed object of
4-7 solar masses (Orosz \& Bailyn 1997; Phillips et al. 1999). 
The compact object is certainly
a black hole, since its mass is beyond the theoretical
maximum mass limit of $\sim$ 3 solar masses for neutron stars 
(Kalogera \& Baym 1996).

King (1998) proposes that the superluminal sources are black hole 
binaries with the secondary in the Hertzsprung-Russell gap, which 
provides super-Eddington accretion into the black hole. In the Galaxy 
there would have 
been $\geq$ 10$^3$ systems of this class with a lifetime for the jet phase 
of $\leq$ 10$^7$ years, which is the spin-down phase of the black hole.

\vsl
\ni 3.1 {\it Superluminal motions in GRS 1915+105}
\vs

Figure 2 shows a pair of bright radio condensations emerging in opposite 
directions from the compact, variable core of GRS 1915+105. 
Before and after the remarkable ejection event shown in Figure 2, the 
source ejected other pairs of condensations but with flux densities one 
to two orders of magnitude weaker. One of these weaker pairs can be seen 
in the first four maps of Figure 2, as a fainter pair of condensations 
moving ahead of the bright ones at about the same proper motion and direction. 
 
In Figure 3 we show the proper motions of the condensations detected 
from four ejection events in 1994. The angular displacements from the 
stationary core are consistent with ballistic (that is, unaccelerated) 
motions. The time separation between ejections suggests a quasi-periodicity 
at intervals in the range of 20-30 days. Although the clouds in each 
event appear to move ballistically, always in the same general region of 
the sky, their position angles suggest changes by $\sim$ 10$^{\circ}$ in the 
direction of ejection in one month. 
%Two additional ejections were followed 
%more than one year later in August 1995 (Mirabel et al. 1996a) and about 
%three years later in October-November 1997 (Fender et al. 1999;
%Dhawan, Mirabel, \& Rodr\'\i guez 1999).  The position angles 
%and proper motions for these later ejections are consistent with those of 
%1994 and 1995 and suggest a precession of $\leq$10$^{\circ}$ in the direction of 
%ejection over intervals of a few years. 
 
Figures 2 and 3 show two asymmetries: one in apparent transverse motions, 
another in brightness. The cloud that appears to move faster also appears 
brighter. It has been shown that both asymmetries, in proper motions and 
in brightness, are consistent with the hypothesis of an anti-parallel 
ejection of twin clouds moving at relativistic velocities 
(Mirabel and Rodr\'\i guez 1994), as discussed in section 4. 
At a distance of 12 kpc the proper motions measured with the VLA in 
1994 of the approaching 
(17.6 $\pm$ 0.4 mas d$^{-1}$) and receding (9.0 $\pm$ 0.1 mas d$^{-1}$) 
condensations shown in Figure 2 imply apparent velocities on the plane of the sky 
of 1.25c and 0.65c, respectively. From the analysis of relativistic distorsion 
effects using the equations in the next section and the
VLA data, it is inferred that the ejecta move with 
a speed of 0.92c at an angle $\theta$ = 70$^{\circ}$ to the line of sight. 

Within the errors of the measurements and a precession of $\leq$10$^{\circ}$, 
relativistic ejections with a stable jet axis at scales of 500-5000 AU 
and larger were later observed from GRS 1915+105 over
a time span of four years (Mirabel et al 1996a; Fender et al 1999; 
Dhawan, Mirabel and Rodr\'\i guez 1999). The VLBA images of GRS~1915+105 
show that the jets are already collimated at milliarcsec scales 
(Dhawan, Mirabel and Rodr\'\i guez  1999), 
namely, at about 10 AU from the compact source (Figure 4). 
The core appears as a synchrotron jet of length $\sim$100 AU 
before and during optically thin flares, and at those scales 
it already exhibits Doppler boosting. Discrete ejecta have appeared 
at about 500 AU. Both, the observations with 
MERLIN (Fender et al. 1999) and with the VLBA (Dhawan, Mirabel and Rodr\'\i guez 1999) 
in the years 1997 and 1998 have shown faster apparent superluminal motions at 
1.3c-1.7c 
at scales of hundreds of AU, and intrinsic expansions of the expelled clouds 
mostly in the direction of their bulk motions. At present it is not clear if 
the faster motions measured with the higher resolution observations of MERLIN and VLBA 
in 1997 relative to the VLA observations in 1994 are due to intrinsic faster 
ejections, changes in the angle to the line of sight, or 
to resolution effects between the arrays as suggested by Fender et al. (1999). 

A secular paralax of 5.8$\pm$1.5 mas yr$^{-1}$ in the galactic plane, in 
rough agreement with the HI distance of 12 kpc (Rodr\'\i guez et al. 1995), 
has been measured with the VLBA 
(Dhawan, Mirabel and Rodr\'\i guez  1999).

\vsl
\ni 3.2 {\it Superluminal motions in GRO J1655-40}
\vs

The relativistic ejections observed in the radio in GRO~J1655--40 have 
striking similarities as well as differences with those in GRS 1915+105. 
Bright components moving apart with proper motions in the range of 
40 to 65 mas d$^{-1}$ were independently observed with the Southern 
Hemisphere VLBI Experiment array (Tingay et al. 1995), and the VLA and VLBA
(Hjellming \& Rupen 1995). In Figure 5 is shown a sequence of seven 
VLBA radio images of GRS J1655-40 from Hjellming \& Rupen (1995). At a 
distance of 3.2 kpc the motions of the ejecta have been fit -using a kinematic 
model- with a velocity of 0.92c, and a jet axis inclined 85$^{\circ}$ to 
the line of sight at a position angle of 47$^{\circ}$, about 
which the jets rotate every three days at an angle of 2$^{\circ}$. 
 
In contrast to what has been observed in the repeated 
ejections of GRS 1915+105, the flux ratios of the blobs on either 
side of GRO J1655--40 cannot 
be ascribed to relativistic Doppler boosting. In GRO J1655--40 the 
asymmetry in brightness appears to flip from side to side
(Hjellming \& Rupen 1995). 
Not only the jets appear to be intrinsically asymmetric, but 
also the sense of that asymmetry changes from event to event. 
Therefore, although similar intrinsic velocities greater than 0.9c are found 
in both superluminal sources, due to the asymmetries in GRO J1655-40, the 
ultimate physical interpretation of the superluminal expansions 
in this source remains uncertain. 

We point ou that in SS433 flux asymmetries between knots ejected simultaneously 
on both sides have also been observed (Fejes, 1986). This asymmetry could be due 
to intrinsic variations, so perhaps GRO J1655-40 is not unusual in this respect. 
However, VLBA multiwavelength monitoring of SS433 (Paragi et al. 1998) shows 
that it is always the receding part of the core-complex which is fainter compared 
to the approaching one, and that this effect cannot be explained simply by 
Doppler beaming. It is possible that free-free absorption and the different 
pathlengths through an absorbing medium could explain some of these asymmetries 
in SS433 and other jet sources. Furthermore, in SS433 more than 90\% of the 
radio emission is in knots rather than in continuous jets, and the core complex 
disappears after large outbursts, as in GRS 1915+105 (Mirabel and Rodr\'\i guez 1994).

\vsl
\ni 3.3 {\it Superluminal motions in XTE J1748-288}
\vs

Two major relativistic ejection sequences moving at least 20 mas/day were observed 
in June 1998 (Hjellming et al. 1998) from the hard X-ray transient XTE J1748-288 
(Smith et al. 1998). Each sequence appeared to begin with a one-sided relativistic 
ejection. The ejecta are highly linearly polarized, and at a distance of 8 kpc, 
derived from the HI $\lambda$21cm absorption line, their motions would imply 
apparent speeds 
of 0.9c and 1.5c, and intrinsic velocities of more than 0.9c (Hjellming et al. 1998). 
This is the first galactic source of relativistic jets where it has been observed 
in real time that the jets collide 
with environmental material, being decelerated while brightening at the leading 
edge of the jet.

\vsl
\ni 4.~SPECIAL RELATIVITY EFFECTS
\vs

\vsl
\ni 4.1 {\it Parameters of the Ejection }
\vs

The main characteristics of the superluminal ejections 
can be understood in terms of the simultaneous
ejection of a pair of twin condensations moving 
at velocity $\beta$ ($\beta = v/c$), with $v$ being the velocity 
of the condensations
and $c$ the speed of light), with the axis of the flow making an
angle $\theta$ ($0^\circ \leq \theta \leq 90^\circ$)
with respect to the line of sight of a distant 
observer (Rees 1966; see Figure 6). The apparent proper motions in the sky
of the  
approaching and receding condensations, $\mu_a$ and $\mu_r$, are given by:
 
$$\mu_a = {\beta~sin~\theta \over {(1 - \beta~cos~\theta)}}
{c \over D}, \eqno(1)$$
 
$$\mu_r = {\beta~sin~\theta \over {(1 + \beta~cos~\theta)}}
{c \over D}, \eqno(2)$$
 
\noindent where D is the distance from the observer to the source.
These two equations can be transformed to the equivalent pair of
equations:
 
$$\beta~cos~\theta = {\mu_a - \mu_r \over \mu_a + \mu_r}, \eqno(3)$$
 
$$D = {c~tan~ \theta \over 2}{(\mu_a - \mu_r) \over \mu_a \mu_r}. \eqno(4)$$
 
If only the proper motions are known, an interesting upper limit for the distance can be
obtained from eqns. (3) and (4):

$$D \leq {c \over {\sqrt {\mu_a \mu_r}}}. \eqno(5)$$

In all equations we use cgs units and the proper motions are in radians s$^{-1}$.
In the case of the bright ejection event of
1994 March 19 for GRS~1915+105,  the proper
motions measured were $\mu_a = 17.6 \pm 0.4$ mas day$^{-1}$ and
$\mu_r = 9.0 \pm 0.1$ mas day$^{-1}$. 
Using eqn. (5), we derive an upper limit for the
distance, $D \leq$ 13.7 kpc, confirming the
galactic nature of the source.
 
The distance to GRS~1915+105 is found to be, from HI absorption studies,
12.5$\pm$1.5 kpc (Rodr\'\i guez et al. 1995; Chaty et al. 1996).
Then, the proper 
motions of the approaching and receding condensations measured with the VLA 
in 1994 and 1995 imply apparent 
velocities on the plane of the sky of $v_a$ = 1.25c and $v_r$ = 0.65c for 
the approaching and receding components respectively. The ejecta move 
with a true speed of $v$ = 0.92c at an angle $\theta$ = 70$^{\circ}$ 
with respect to 
the line of sight (Mirabel \& Rodr\'\i guez 1994). The faster proper 
motions of 24 mas/day measured with MERLIN (Fender et al. 1999) and 
the VLBA (Dhawan et al. 1999) in 1997 would imply a true speed of 0.98c 
at an angle of 66$^\circ$ to the line of sight.

\vsl
\ni 4.2 {\it A Relativistic Distance Determination }
\vs
 
We note that the detection of a known line from either of the condensations
would allow a precise determination of the distance.
The Doppler factors, namely, the ratios of observed to emitted frequency
($\nu_0$) for the approaching and receding condensations are given
by
 
$$ \delta_a = {\nu_a \over \nu_o} = \gamma^{-1} (1 - \beta~cos~ \theta)^{-1},
\eqno(6) $$
 
$$ \delta_r = {\nu_r \over \nu_o} = \gamma^{-1} (1 + \beta~cos~ \theta)^{-1}
 \eqno(7) $$
 
\noindent In these
last two equations $\gamma = (1 - \beta^2)^{-1/2}$ is the Lorentz factor.
Since we know $\beta~cos~ \theta$, a determination of either
$ {\nu_a / \nu_o}$ or $ {\nu_r / \nu_o}$ will allow the determination
of $\beta$ and thus the determination of $\theta$ and of the
distance from eqn. (4).
In the case of cosmologically distant objects, the equations 1, 2, 4, and 5 
are valid replacing the distance $D$ by the angular size
distance $D_a$ (Peebles 1993), and
the rest frequency $\nu_o$ by $\nu_o /(1+z)$, with $z$ being the
observed redshift of the central source. 
The angular size distance is given by
$D_a = (c~z/H_0)[1 - (1 + q_0) z/2 + ...]$, where
$H_0$ is Hubble's constant and $q_0$ is the dimensionless
acceleration (or deceleration) parameter.
Then, the observations 
of proper motions and frequency shifts in extragalactic
relativistic ejecta pairs could 
potentially be used to test between different cosmological
models. 
 
\vsl
\ni 4.3 {\it Doppler Boosting }
\vs
 
The ratios of observed to emitted flux density $S_o$, from a twin pair of
optically-thin, isotropically emitting jets are:
 
$$ {S_a \over S_o} = \delta_a^{k-\alpha}, \eqno(8) $$
 
$$ {S_r \over S_o} = \delta_r^{k-\alpha}, \eqno(9) $$
 
\noindent where $\alpha$ is the spectral index of the emission
($S_\nu \propto \nu^\alpha)$, and $k$ is a parameter that
accounts for
the geometry of the ejecta, with k = 2 for a continuous jet
and k = 3 for discrete condensations.
Then, the ratio of observed flux densities (measured at equal
separations from the core) will be given by
 
$$ {S_a \over S_r} = \biggl({1 + \beta~cos~ \theta
\over 1 - \beta~cos~ \theta} \biggr)^{k-\alpha}, \eqno(10)$$
 
Since for the 1994 March 19 event
$\beta~cos~\theta$ = 0.323 and $\alpha$ = - 0.8 the flux ratio in the case of 
discrete condensations would be 12, whereas for a continuous jet it would be 6.
For a given angular separation it was found that the observed flux ratio between 
the approaching and receding condensations is 8 $\pm$ 1. Similar results were 
found using the MERLIN observations by Fender et al. (1999). 
Therefore, irrespective of the distance to the source, the flux ratios for equal
angular separations from the core are consistent with the assumption of a twin 
ejection at relativistic velocities.
Atoyan \& Aharonian (1997) have considered the observable effects
in the flux density ratio
of asymmetries between the jet and counterjet.
Bodo \& Ghisellini (1995) have proposed that there could be a contribution
of wave propagation in the pattern motions, but that most of the observed displacements
are true bulk plasma velocities.

\vsl
\ni 5. ACCRETION DISK INSTABILITIES AND JET FORMATION
\vsl
 
Collimated jets seem to be systematically associated with
the presence of an accretion disk around a star or a collapsed object.
In the case of black holes,
the characteristic dynamical times in the flow of matter 
are proportional to the black hole's mass, and the events with 
intervals of minutes in a microquasar could correspond
to analogous phenomena with duration of thousands of years in a quasar of 
10$^9$ M$_{\odot}$ (Sams et al. 1996; Rees, 1998).
Therefore, the variations with minutes of duration observed in a microquasar  
in the radio, IR, optical, and X rays could 
sample phenomena that we have not been able to observe in quasars.
 
X-rays probe the inner accretion disk region, radio waves the synchrotron 
emission from the relativistic jets. The long term multiwavelength 
light curves of the superluminal sources show that the hard X-ray emission 
is a necessary but not sufficient condition for the formation of collimated 
jets of synchrotron radio emission. In GRS 1915+105 the relativistic ejection 
of pairs of plasma clouds have always been preceded by unusual activity 
in the hard X-rays (Harmon et al. 1997),
more specifically, the onset of major 
ejection events seems to be simultaneous to the sudden drop from a 
luminous state in the hard X-rays (Foster et al. 1996;
Mirabel et al. 1996a). However, not all unusual 
activity and sudden drops in the hard X-ray flux appear to be associated 
with radio emission from relativistic jets. In fact, in GRO J1655-40 there have 
been several hard X-ray outbursts without following radio flare/ejection events. 
A more detailed summary of the long term multifrequency studies of black 
hole binaries can be found in Zhang et al. (1997).
 
The episodes of large amplitude X-ray flux variations in time-scales of 
seconds and minutes, and in particular, the abrupt dips observed
(Greiner et al. 1996; Belloni et al. 1997; Chen et al. 1997) in 
GRS~1915+105 are believed to be evidence for the presence of a black hole,
as discussed below. 
These variations could be explained if the inner ($\leq$ 200 km) part of the 
accretion disk goes temporarily into an advection-dominated mode
(Abramowicz et al. 1995; Narayan et al. 1997). In 
this mode, the time for the energy transfer from ions (that get most of the 
energy from viscosity) to electrons (that are responsible for the radiation) 
is larger than the time of infall to the compact object. Then, the bulk 
of the energy produced by viscous dissipation in the disk is not radiated 
(as it happens in standard 
disk models), but instead is stored in the gas as thermal energy.
This gas, with large amounts of locked energy, is advected (transported)
to the compact object. If the compact object is a black hole, the energy
quietly disappears through the horizon. In constrast, if the compact object is a
neutron star, the thermal energy in the superheated gas 
is released as radiation when it collides with the surface of the neutron 
star and heats it up. The cooling time of the neutron star photosphere 
is relatively long, and in this case a slow decay in the X-ray flux is 
observed. Thus, one would expect the luminosity of black hole binaries
to vary over a much wider range than that of neutron star
binaries (Barret et al. 1996). The idea of advection-dominated flow has 
also been proposed (Hameury
et al. 1997) 
to explain the X-ray delay in an optical outburst (Orosz
et al. 1997) of GRO J1655-40.

During large-amplitude variations in the X-ray flux of GRS 1915+105, 
remarkable flux variations on time-scales of minutes have also been reported 
at radio (Pooley \& Fender 1997; Rodr\'\i guez \& Mirabel 1997; Mirabel et al. 1998) 
and near-infrared wavelengths (Fender et al. 1997; Fender and Pooley, 1998; 
Eikenberry et al. 1998a; Mirabel et al. 1998).  
The rapid flares at radio and infrared waves are thought to come 
from expanding magnetized clouds of relativistic particles. 
This idea is supported by the observed time shift of the emission at radio waves 
as a function of wavelength and the finding of infrared synchrotron precursors 
to the follow-up radio flares (Mirabel et al. 1998). Sometimes the 
oscillations at radio waves appear as isolated events composed of twin flares 
with characteristic time shifts of 70$\pm$20 minutes 
(e.g. Pooley \& Fender, 1997: Dhawan, Mirabel \& Rodr\'\i guez, 1999).
The time shift between the twin peaks seems to be 
independent of wavelength (Mirabel et al. 1998), and no Doppler boosting 
is observed. This suggests that these quasiperiodic flares may come from expanding 
clouds moving in opposite directions with non-relativistic bulk motions. 
 
In Figure 7 are shown simultaneous light curves in the X-rays, infrared, 
and radio wavelengths, together with the X-ray photon index during a large 
amplitude oscillation. These light curves can be consistently
interpreted to imply  that the relativistic clouds of plasma 
emerge at the time of the dips and follow-up recovery of the X-ray flux. 
In adiabatically expanding clouds the maximum flux density at short 
wavelengths (i.e. the near infrared) should be observed very shortly after 
the ejection (10$^{-3}$ sec), and it is only in the radio wavelengths that 
significant time delays occur (Mirabel et al. 1998). Figure 7 shows that 
the onset of the infrared flare occured $\geq$ 200 sec after the drop of 
the X-ray flux, during its recovery from the dip, probably
at the time of
the appearance of an 
X-ray spike (t = 13 min) which is associated to a sudden softening of 
the (13-60 keV)/2-13 keV) photon index 
due to the drop in the hard X-ray flux. 
Similar phenomena have been observed in this source
by Eikenberry et al. (1998a). In the context of the unstable 
accretion disk model of Belloni et al. (1997), these observations 
suggest that the ejection of plasma clouds takes place during the 
subsequent replenishment of the inner accretion disk,
well after the disappearance of the soft component at the sudden
drop. The ejection of the clouds seems to be coincident with the
soft X-ray peak at the dip.  
Furthermore, the slow rise of the infrared flux to maximum seen in Figure 7 
indicates that the injection of relativistic particles is not instantaneous 
and that it could last up to tens of minutes.        
 
Mirabel et al. (1998) have estimated that the minimum mass of the clouds 
that are ejected every few tens of minutes is $\sim$ 10$^{19}$ g. 
% (about the mass of Mount Everest). 
On the other hand, the estimated total 
mass that is removed from the inner
accretion disk in one cycle of a few tens of minutes is of the order of 
$\sim$10$^{21}$ g (Belloni et al. 1997). Given the 
uncertainties in the estimation of these masses, it is still unclear 
what is the fraction of mass of the inner accretion disk that disappears through 
the horizon of the black hole. Anyway, it seems plausible that during 
accretion disk instabilities consisting on the sudden disappearance 
of its inner part, most of it is advected into the black hole, and only some 
fraction is propelled into synchrotron-emitting clouds of plasma.  
 
Energy outbursts in the flat synchrotron spectrum over at least four decades 
of frequency have also been observed in Cygnus X-3 
(Fender et al. 1996). The optical polarization 
observed in GRO J1655-40 (Scaltriti et al. 1997) could also be 
related to the presence of synchrotron emission at optical wavelengths.
The study of GRS 1915+105 lead to the realization that besides the energy 
invested in the acceleration of the plasma clouds to their bulk motions, the 
oscillations of the type shown in Figure 7 require synchrotron
luminosities of at least 
10$^{36}$ erg s$^{-1}$.
This synchrotron  luminosity is not 
negligible with respect to the thermal luminosity radiated in the X-rays. These 
results give support to the observation of 
synchrotron infrared jets reaching distances 
of a few thousand AU from GRS 1915+105 (Sams et al. 1996).

\vsl
\ni 6.~JET FORMATION
\vsl
 
The processes by which the jets are accelerated and collimated
are still not clearly understood, but it is
believed that several of the concepts proposed for extragalactic
jets can be extended to galactic jets.

Blandford \& Znajek (1977) take advantage of the fact that, in principle,
it is possible to extract energy and angular momentum from
a rotating black hole (Penrose 1969), to produce electric and magnetic fields
and possibly fast outflowing jets. 
A magnetized accretion disk around the Kerr black hole
brakes it electromagnetically. However, Ghosh and Abramowicz (1997) and 
Livio et al. (1998) have called into question that the Blandford-Znajek 
process can provide the primary power in the jets.

A seminal idea that has been followed by many researchers in the
field is that of
the magnetohydrodynamical model of Blandford \& Payne (1982).
These authors proposed that the angular momentum of a magnetized
accretion disk around the collapsed object is the responsible for
the acceleration of the plasma. The magnetic field lines are
taken to be frozen into
the disk and the plasma is assumed to follow them
like a ``bead on a wire'', at least close to
the disk. If the field line forms an angle with the plane of the disk
smaller than 60$^\circ$, the displacements of the
plasma from its equilibrium position become unstable.
This happens because
along these field lines the component of the centrifugal
force will be larger than the component of the
gravitational force and the plasma will
be accelerated outwards.
Then, in its origin, the outflow motion has
an important ``equatorial'' component, 
while on larger scales the
jets are observed to
have a motion that is dominantly ``poloidal''.
In other words, after the acceleration
a collimating mechanism is required
to change the wide-angle centrifugal outflow into a collimated jet.

This collimation is proposed to be achieved as follows.
Inside an inner region, the magnetic field energy 
density is larger than 
the kinetic energy density of the flow but at some distance from
the disk (the Alfv\'en surface), this situation 
reverses and the flow stops corotating with
the disk. This causes that a loop of toroidal
(azimuthal) field is added to the flow for each rotation of
the footpoint of the field line.
The tension of this wound-up toroidal field that is formed 
external to the
Alfv\'en surface produces a force directed toward the axis
(the ``hoop stress'') that eventually collimates the flow
into a jet. 
Most models for the production of jets in the astrophysical 
context use elements of MHD acceleration and collimation. 

Recently, several groups (Spruit, Foglizzo, \& Stehle 1997;
Lucek \& Bell 1997;
Begelman 1998) have pointed out that the toroidal
field traditionally held responsible for collimating
jets in the MHD mechanism is
unstable and cannot collimate the jets effectively.
It has been proposed alternatively that the collimating agent
is the poloidal component of
the magnetic field. 

Koide et al. (1998) have performed for the
first time full general relativistic MHD numerical simulations
of the formation of jets near a black hole. Their results
suggest that the ejected jet has a two-layer 
structure with an inner, fast gas-pressure driven component and
an outer, slow magnetically-driven component.
The presence of the inner, fast gas-pressure driven component
is a result of the strong pressure increase produced by shocks 
in the disk through fast advection flows inside the
last stable orbit around a black hole. This feature is not
seen in non-relativistic calculations.

Within the uncertainties of the small sample,
the velocity of the jets seems to show a bimodal distribution,
with some sources having $v_{jet} \simeq$ 0.3c
and others having $v_{jet} \geq$ 0.9c.
Two explanations have been offered in the literature.
On one hand, Kudoh \& Shibata (1995) suggest that
the terminal velocity of the jet is of order of the Keplerian
velocity at the footpoint of the jets, that is that
the fastest jets probably come from the deepest
gravitational wells (Livio, 1998).
On the other hand, Meier et al. (1997) propose that the
velocity of the jets is regulated by a magnetic ``switch'', with
highly relativistic velocities achieved only above a critical
value of the ratio of the Alfv\'en velocity to the escape velocity.
The determination of the mass of the collapsed object in the
jet sources would discriminate between these two models.

While it seems that a steady state MHD model can account for the
formation of continuous relativistic jets, the
events discussed by Mirabel et al. (1998), Belloni
et al. (1998), and Fender \& Pooley (1998)    
that seem to involve a connection between the disappearance of the inner
accretion disk and the sudden ejection of condensations may require a different
mechanism. 
%In the MHD models the disk is the launching 
%platform of the jets and cannot ``eject'' itself.
Clearly, the time seems to be ripe for
new theoretical advances
on the models of formation of relativistic jets
that take into account the observational
features found in stellar jets.

Another characteristic that the jet models must account
for is the production of relativistic particles that
will produce the synchrotron emission that is
observed in several sources.
As in other astrophysical contexts, it is believed
that the acceleration of electrons to relativistic speeds takes place in
shocks (Blandford \& Ostriker 1978).
On the other hand,
most of the X-ray binaries are ``radio-quiet'', implying that
relativistic electrons and/or magnetic fields are not
always present in sufficient amounts. 
% These ``radio-quiet''
% X-ray binaries could, however, have jets that
% emit thermally (and are detectable only in the X-rays).

\vsl
\ni 7.~SYNCHROTRON EMISSION
\vs

The high brigthness temperature, rapid variability, and linear
polarization observed in the radio emission from X-ray binaries
indicates a synchrotron origin.
The time evolution of the radio emission has been modeled in terms
of conical jets or expanding clouds of magnetized plasma
(Hjellming \& Johnston 1988; Mart\'\i\ et al. 1992; Seaquist 1993).

In the simplest case of an adiabatically expanding spherical cloud in 
the optically-thin regime,
the van der Laan (1966) model is used, where 
the flux density is given by
$ S_\nu \propto \nu^{(1-p)/2} ~ r^{-2p}$,
and the relativistic electrons have an energy distribution
given by $N(E) = KE^{-p}$, with $K$ being a constant that
is related to the density of the relativistic electrons.
In this equation $r$ is the radius of the cloud.
Assuming that the cloud expands linearly with time,
% as 
%a power law, $r \propto t^{a}$ with $t$ being the time
%since the start of the expansion, the 
the flux density is
given by
$ S_\nu \propto \nu^{(1-p)/2} ~ r^{-2p}$. 
Assuming a typical value of $p$ = 2.4, one
% and that the cloud expands linearly (a=1), one 
obtains
$ S_\nu \propto \nu^{-0.7} ~ t^{-4.8}$. 
This simple model fits the flux decrease reasonably
well for several of the radio-emitting X-ray binaries (Ball 1996).
However, in some of the best studied
jet sources (SS~443, Hjellming \& Johnston 1988, Vermeulen et al 1993;
GRS~1915+105, Rodr\'\i guez \& Mirabel 1999a),
much less steep decreases are observed.
This situation
can be accounted for by making modifications to the
simple expanding model. One possibility is to
attribute this shallower drop
of flux density with time to constrained
expansion (the source cannot expand
in 3 dimensions but only in 1 or 2 
dimensions). In fact, the GRS 1915+105 maps with milliarcsec resolution 
by Dhawan et al. (1999) show that the expansion of the clouds at hundreds of 
AU from the compact source is mostly in one direction. 
% or to a non-lineal expansion with time
%($a \leq$ 1). 
The flux density can be then
approximately described as
$ S_\nu \propto \nu^{-0.7} ~ t^{-(2/3) p n}$,
where $n$ is the number of dimensions where
expansion is allowed.
Both in SS~433 (Hjellming \& Johnson 1988) and in
GRS~1915+105 (Rodr\'\i guez \& Mirabel 1999a), a break in the
power law that describes the decrease in flux as a function of
time is observed. Remarkably, in both sources the decrease
close to the source can be described with $S_\nu \propto t^{-1.3}$,
while after a distance of $\sim 2 \times 10^{17}$ cm, $S_\nu \propto t^{-2.6}$
is observed. Hjellming \& Johnson (1988) have proposed that
these power laws can be explained as a result of an initial slowed expansion
followed by free expansion in two dimensions. 
This steepening of the decrease in flux density
with angular separation
could be related to the similar tendency observed in the jets of some
radio galaxies, where the intensity $I$ declines with angular
distance $\phi$ as $I_\nu \propto \phi^{-x}$,
with $x$ = 1.2-1.6 in the inner regions and 
$x \sim$4 in the outer regions of the jet (Bridle \& Perley 1984). 

It is also possible that continued injection of relativistic
particles and/or magnetic field into
the emitting plasma can produce shallower decreases with
time of the flux density (Mirabel et al. 1998).
The particle injection could result from in situ acceleration
as the moving gas shocks and entrains ambient gas or
could result from beams or winds from the central energy source.
%Finally, the decline of optically-thin, adiabatically expanding conical
%jets goes as $S_\nu \propto ~ r^{-7(p-1)/6}$,
%that is, is much less steep than that of discrete clouds.
%Hjellming \& Johnston (1988) have modeled the brightness as a function of
%distance of the radio jet of SS~433 in terms of a biconical jet
%that undergoes slowed expansion close to the source and later on
%has a transition to free expansion.
%The different decay behaviors for the radio emission from
%X-ray sources may be related to the detectability of jets.
%Sources decaying rapidly as predicted on the simple adiabatic model
%are very difficult to detect after some time, while those
%sources with slower decay can be followed in time and the jets and
%their motion detected.
The optically thick rise occurs very rapidly and has yet to be observed in detail
for a proper comparison with the theoretical expectations.

It is possible to estimate the parameters of the ejected condensations
using the formulation of Pacholczyk (1970) for minimum energy,
correcting for relativistic effects and 
integrating the radio luminosity over the observed range of frequencies.
Rodr\'\i guez \& Mirabel (1999a) estimate for the bright
1994 March 19 event in GRS 1915+105 a magnetic field of 
about 50 mGauss
and an energy of about $4 \times 10^{43}$ ergs in the relativistic
electrons. Assuming that there is one (non-relativistic) proton per 
(relativistic) electron one gets a proton mass estimate in the order 
of 10$^{23}$ g. To estimate the peak mechanical power
during the ejection we need a value for the time
over which the acceleration and ejection took place.
Mirabel \& Rodr\'\i guez (1994) conservatively estimate that the
ejection event must have lasted $\leq$ 3 days, requiring
a minimum power of $\sim 5 \times 10^{38}$ erg s$^{-1}$, a value comparable with 
the maximum observed steady photon luminosity of GRS~1915+105,
which is $\sim 3 \times 10^{38}$ erg s$^{-1}$ (Harmon et al. 1994).

The ejection events that preceded and followed
the 1994 March 19 outburst are estimated to have masses in
the order of 10$^{21-22}$ g. Finally, if the 
repetitive events observed with periods
of tens of minutes in GRS~1915+105 
(Rodr\'\i guez \& Mirabel 1997; Pooley \& Fender 1997;
Mirabel et al. 1998; Eikenberry et al. 1998a) are interpreted as mini-ejection
episodes, the mass associated with them
is of order 10$^{19}$ g. We crudely estimate that, on
the average, GRS~1915+105 injects energy in the order of 10$^{23}$ g
per year in the form of relativistic (0.92c0.98c), collimated outflows. 
This corresponds to an average mechanical energy of
$L_{mech} \sim 10^3 L_\odot$. In contrast, SS~433 as a result
of its more continuous jet flow, has $L_{mech} \sim 10^5 L_\odot$
(Margon 1984)
despite having a lower flow velocity than GRS~1915+105.
The GRS~1915+105 bursts are thus very energetic but more sporadic.

Recently, there has been evidence that during
some events the synchrotron emission
in GRS~1915+105 extends from the radio into at least
the near-infrared
(Mirabel et al. 1998; Fender \& Pooley 1997). Then
the synchrotron luminosity becomes significant,
reaching values of $ 10^{36}$ erg s$^{-1}$.

As emphasized by Hjellming \& Han (1995), relativistic plasmas are 
difficult to confine and synchrotron radiation sources in stellar environments
will tend to be variable in time. Then, one of the behaviors most difficult 
to account for is the relative constancy of the radio flux in some sources, 
of which Cyg X-1 is the extreme example. The presence of a steady outflow 
that is too faint to be followed up in time as synchrotron-emitting
ejecta could be
consistent with the lack of large variability in this type of source.

\vsl
\ni 8. POSSIBLE LABORATORIES FOR GENERAL RELATIVITY
\vsl

The X-ray power of the superluminal sources  exhibits a large variety of 
quasi-periodic oscillations (QPOs) of high frequency. Of particular interest 
is the class of fast oscillations 
with a maximum stable frequency of 67 Hz 
observed many times in GRS 1915+105, irrespective of the X-ray
luminosity of the source (Morgan et al. 1997).
A QPO with maximum fix frequency of 300 Hz has been observed in GRO J1655-40 
(Remillard et al. 1998). 
These stable maximum frequencies are not seen at times of strong radio flares 
or jet injection. They are believed to be a function of the fundamental properties of the 
black holes, namely, their mass and spin. 
 
One possible interpretation is that these 
frequencies correspond to the last stable circular 
orbit around the black hole. This frequency depends 
on the black hole's mass and spin, as well
as on the rotation direction of the accretion disk, and offers the prospect 
of inferring the spin of black holes with masses independently determined. 
Since from optical observations the mass of the hole in GRO J1655-40 is 
known to be in the range of 4-7 solar masses, one can conclude that 
GRO J1655-40 contains a Kerr black hole rotating at $\geq$ 70\% 
of the maximum spin possible (Zhang et al. 1997). 
 
Alternatively, the maximum QPO stable frequency could be related to 
general relativity disk seismology, more specifically, to the maximum radial 
epicyclic frequency (Nowak et al. 1997), which also depends on the spin of the black hole. 
 
A third interpretation has been proposed in terms of 
the relativistic dragging of the inertial frame around the spinning 
black hole (Cui et al. 1998). 
By comparing the computed disk precession frequency with that of the QPO, the 
spin can be derived if the mass is known. The two sources of sporadic 
superluminal jets are found to be the black holes that spin at rates close 
to maximum limit. Obviously, theoretical work to distinguish between these 
three alternative interpretations will be important to estimate the spin 
of the black holes with known masses.

X-ray spectroscopy of the two superluminal sources obtained with the satellite 
ASCA (Ebisawa 1996; Ueda, 1998) has shown K$_{\alpha}$ H and He like iron absorption 
lines, whereas the observations with SAX have only shown emission features from 
the relativistic accretion disk around 7 keV, which 
have been interpreted as iron lines (Matt et al. 1998). One expects that with 
greater sensitivity 
these lines will show a profile reminiscent of that of the asymmetric iron lines 
observed in Seyfert galaxies (Tanaka et al. 1995). The accretion disks of GRS 1915+105 
and GRO J1655-40 are viewed obliquely, and the blueshifted side of the lines 
should  look much stronger due to the Doppler beaming effect. In addition, the 
center of the line should be redshifted as expected from general relativity 
effects on radiation escaping from the surroundings of a strongly 
gravitating object. In the future, perhaps these lines 
could be used as probes of general relativity effects in 
the innermost parts of the accretion flows into black holes.

General relativity theory 
in weak gravitational fields has been successfully 
tested by observing in the radio
the expected decay in the orbit of a 
binary pulsar, an effect produced by gravitational radiation
damping (Taylor \& Weisberg 1982). Observations of 
binary pulsars have also been used to constrain
the nature of gravity in the strong-field regime (Taylor et al. 1992). 
Although the interpretation of the maximum stable 
frequency of the X-ray power spectrum
in the superluminal sources is still uncertain, 
these frequencies are known to originate close to the horizon 
of the black hole,  and perhaps they could be used in the future to test the physics
of accretion disks and black holes in the strong field limit.

\vsl
\ni 9. OTHER SOURCES OF RELATIVISTIC JETS IN THE GALAXY 
\vsl

In X-ray binaries there is a general correlation between the X-ray 
properties and the jet properties. The time interval and flux amplitude of the 
variations in radio waves seems to correspond to the time and amplitude 
variations in the X-ray flux. More specifically, persistent X-ray sources 
are also persistent radio sources, and the transient X-ray sources 
produce at radio waves sporadic outburst/ejection events.  
Persistent sources of hard X-rays (e.g. 1E1740.7-2942, GRS 1758-258) are 
usually associated to faint, double-sided radio structures that have sizes 
of several arcmin (parsec scales). The radio core of these two persistent sources 
are weak ($\leq$ 1 mJy) and do not exhibit high amplitude variability. On the contrary, 
rapidly variable hard X-ray transients (e.g. GRS 1915+105, GRO J1655-40, XTE J1748-288) 
may exhibit variations in the X-rays and radio fluxes of several orders 
of magnitude in short intervals of time. Because these black-hole X-ray 
transients produce sporadic ejections of discrete, bright plasma clouds, the 
proper motions of the ejecta can be measured.

Conservation of angular momentum in accretion disks indicates that probably 
all hard X-ray sources that accrete at super-Eddington rates must produce 
relativistic jets. However, the observational study of these jets presents in 
practice several difficulties. Persistent hard X-ray sources like 
Cygnus X-1 are surrounded by faint non-thermal 
radio features extending several 
arcmin (Mart\'\i\ et al. 1996), and even in the cases where
they are well aligned 
with the variable compact radio counterpart it is very difficult
to prove conclusively that the faint and extended radio features are actually 
associated with the X-ray source. 
This was the case of Sco X-1, where possible large-scale
radio ``lobes'' were found to be extragalactic
sources symmetrically located in the plane of the sky
with respect to Sco X-1 (Fomalont \& Geldzahler 1991).
On the other hand, in transient black hole binaries one may observe 
transient sub-arcsec jets, but unless the interferometric observations 
are conveniently scheduled, the evolution is too rapid and it may not 
be possible to follow up the proper motions of discrete clouds. This may have 
been the case in the radio observations of the X-ray sources Nova Oph 93 
(Dela Valle, Mirabel, \& Rodr\'\i guez  1994) and Nova Muscae (Ball et al. 1995), 
among others.   
 
We list in Table 1 the 10 sources of relativistic jets in the Galaxy known 
so far. The first six are transients, whereas the next three are persistent 
X-ray sources. Proper motions of the relativistic ejecta have been determined 
with accuracy in GRS 1915+105, GRO J1655-40, XTE J1748-288, and 
SS 433. Besides these four sources, proper motions were also measured 
-but with less accuracy- for moving features in Cygnus X-3 
(Schalinski et al. 1995, Mart\'\i\ et al. 1999), Circinus X-1 (Fender et al. 1998), and 
CI Cam (XTE J0421+560; Hjellming \& Mioduszewski 1998; Mioduszewski et al. 1998).
It is interesting that the ejecta from the black hole binaries GRS 1915+105, 
GRO J1655-40, and probably also XTE J1748-288  have velocities greater than 0.9c, 
while the ejecta from the four sources believed to be neutron star binaries 
have velocities $\leq$0.3c. From their
models of magnetically driven jets,
Kudoh \& Shibata (1995) have proposed
that jet velocities such as those listed in Table 1 are comparable 
to the Keplerian rotational velocities expected at the base of the jets, close 
to neutron stars and black holes,  respectively. 
Livio (1998) has also stressed the similarity between the
velocity of jets and the escape velocity of the gravitational well
from where they were ejected.
If this notion is confirmed, jet velocities could then be used to 
discriminate between neutron stars and black holes, with
jet velocities close to the speed of light been produced only
in black hole binaries.

Another possible source of relativistic jets in the Galaxy is, of course, 
Sgr A*, the presumed black hole of 2.5 million solar masses at the galactic 
center (Eckart \& Genzel 1997). The radio source is always present at
about the 1 Jy level and exhibits 
a flat spectrum with relative small variations, a behavior similar
to that of the faint compact mJy radio sources 
associated with Cygnus X-1 (Mart\'\i\ et al. 1996) and GRS 1915+105 in 
its plateau state at times when no strong outburst/ejection events take place,
a state that in the latter source 
can last from days to weeks (Pooley \& Fender 1997). This 
type of radio emission could arise from a jet in a coupled 
jet-disk system (Falcke et al. 1993) or 
from electrons in an advection dominated flow (Narayan et al. 1998; 
Mahadevan 1998; Begelman and Blandford, 1999). Despite heavy interstellar 
scattering at radio 
wavelengths, recent VLBA observations at 7-mm
may have resolved Sgr A* in an elongated radio 
source of 72 Schwarzschild radii suggesting the presence of a jet (Lo et al.
1998).

\vsl
\ni 10. INTERACTION OF RELATIVISTIC JETS WITH THE ENVIRONMENT
\vsl

If a compact source (black hole or neutron star) injects 
% X-ray photons and
collimated relativistic jets into its cold environment, it is expected
that some fraction of the injected power will be dissipated by shocks in the
circumstellar gas and dust. The collision of relativistic ejecta with environmental 
material has been observed in real time in XTE J1748-288 (Hjellming et al. 1998), where 
the leading edge of the jet decelerates while strongly brightening. The interaction 
of the mildly relativistic jets from CI Cam (Hjellming \& Mioduszewski, 1998) with an HII 
and dust shell nebula has been reported by Garc\'\i a et al. (1998). 
Other signatures of the interaction of relativistic jets with the environment are 
the radio lobes of 1E 1740.7-2942 (Mirabel et al. 1992) and
GRS 1758-258 (Rodr\'\i guez et al. 1992), the twisted arcmin jets of
Circinus X-1 (Steward et al. 1993; Fender et al. 1998), and the two lateral 
extensions of tens of
pc in the radio shell W50 that hosts at
its center SS 433. The interaction of SS 433 with the nebula 
W50 has been studied in the
X-rays (Brinkmann et al. 1996), infrared (Mirabel et al. 1996b),
and radio wavelengths (Dubner et al. 1998 and references therein).
 
SS 433 is a high mass X-ray binary at a
distance of $\sim$ 3 kpc near the centre of the radio shell W50
(Margon, 1984). The latter may be either the supernova remnant from the formation
of the compact object (Velusamy \& Kundu 1974), or a bubble evacuated by the
energy outflow of SS 433 (Begelman et al. 1980). Besides the well known
relativistic jets seen at sub-arcsec scales in the radio, large-scale
jets become visible in the X-rays
at distances $\sim$ 30 arcmin ($\sim$ 25 pc) from the
compact source (Brinkmann et al. 1996). In the radio,
the lobes reach distances of up to 1$^\circ$ ($\sim$50 pc).
These large-scale X-ray jets and radio lobes
are the result of the interaction of the mass outflow with the
interstellar medium. From optical and X-ray emission lines it is found that
the sub-arcsec relativistic jets have a kinetic energy of $\sim$ 10$^{39}$ erg s$^{-1}$
(Margon, 1984; Spencer, 1984), which is several orders of magnitude larger than the energy
radiated in the X-rays and in the radio. 
 
In Figure 8 is shown the $\lambda$20 cm map with 55 arcsec resolution by Dubner
et al. (1998).
% , combined with the ROSAT X-ray map by Brinkmann et al. (1996). 
It shows the connection between the subarcsec relativistic jets and the extended 
nebula over $\sim$ 10$^5$ orders of magnitude in distance scales. Dubner et al. (1998)
estimate that the kinetic energy transferred into the ambient medium is
$\sim$ 2 10$^{51}$ ergs, thus confirming that the relativistic jets from SS433
represent an important contribution to the overall energy budget of 
the surrounding nebula W50.
Begelman et al. (1980) characterized W50 as a ``beambag", interpreting the
elongated shape and filled-in radio structure of W50 as evidence for continuing
injection of magnetic field and high-energy particles from SS433.
 
Evidences for the interaction of jets with the environmental medium have
also been searched in the two superluminal sources. In GRS 1915+105
Chaty,
Rodr\'\i guez \& Mirabel (1999) searched at millimeter, infrared,
and X-rays for evidences of the physical association between the relativistic
jets and two IRAS sources projected symmetrically on each side at $\sim$15 arcmin 
of angular distance from the 
compact source that
at first glance could be lobes caused by the impact of the jets
in interstellar molecular clouds (Rodr\'\i guez \& Mirabel, 1999b).
Besides the good alignment of the IRAS sources with the subarcsec jets
and the presence of an intriguing non-thermal jet-like source
in the SE IRAS source
(Rodr\'\i guez \& Mirabel, 1999b), no conclusive
physical evidence for association with 
GRS 1915+105 has been found, with the IRAS sources most probably being
normal HII regions.
On the other hand, Hunstead et al. (1998) find regions of extended
low-surface-brightness emission aligned with the radio jets of GRO J1655-40,
but their real association with the high energy source have not been confirmed.
The jets in GRS~1915+105 and GRO J1655-40 are faster than those in SS~433,
but much more sporadic, and this probably accounts for the lack of
obvious lobes associated with them.
 
It has been proposed that the interaction of relativistic jets with the environment 
may induce high energy radiation. Positrons released impulsively from the compact 
source could annihilate locally in the hot plasma producing a broad 511 keV 
spectral feature (Sunyaev et al. 1991; Ramaty et al. 1992). 
Alternatively, a fraction of the positrons 
could stream up to the interstellar gaseous environment, slowing down and 
annihilating in such cold medium, thus emitting 511 keV narrow line emission, 
inducing radio lobe synchrotron emission and bremsstrahlung gamma-ray continuum emission 
(Laurent \& Paul 1994). 

\vsl
\ni 11. MICROBLAZARS AND GAMMA-RAY BURSTS 
\vsl

It is interesting that in all three sources 
where $\theta$ (the angle 
between the line of sight and the axis of ejection) has been determined,
a large value is found (that is, the axis of ejection is close to the
plane of the sky). These values are
$\theta \simeq 79^\circ$ (SS~433; Margon 1984),
$\theta \simeq 66^\circ-70^\circ$ (GRS~1915+105; Mirabel
\& Rodr\'\i guez 1994; Fender et al. 1999), $\theta \simeq 85^\circ$
(GRO J1655-40; Hjellming \& Rupen 1995), and $\theta \geq 70^\circ$ 
for the remaining sources. This result is not inconsistent with
the statistical expectation since the probability of
finding a source with a given $\theta$ 
is proportional to $sin~\theta$. We then expect to find as many
objects in the $60^\circ \leq \theta \leq 90^\circ$ range
as in the $0^\circ \leq \theta \leq 60^\circ$ range.
However, this argument suggests that we should eventually detect
objects with a small $\theta$. For objects with $\theta \leq 10^\circ$
we expect the timescales to be shortened by
2$\gamma$ and the flux densities to be boosted by 8$\gamma^3$ with respect to  
the values in the rest frame of the condensation.
For instance, for motions with $v$ = 0.98c ($\gamma$ = 5), the timescale will shorten
by a factor of $\sim$10 and the flux densities will be boosted by
a factor of $\sim 10^3$. Then, for a galactic source with
relativistic jets and small $\theta$ we expect fast and intense
variations in the observed flux.
These microblazars may be quite hard to detect
in practice, both because of the low probability
of small $\theta$ values and because of the fast decline
in the flux.

Gamma-ray bursts are at cosmological distances and ultra-relativistic 
bulk motion and beaming appear as essential ingredients to 
solve the enormous energy requirements. Beaming reduces 
the energy release by the beaming factor f = $\Delta$$\Omega$/4$\pi$, where 
$\Delta$$\Omega$ is the solid angle of the beamed emission. 
Additionally, the photon energies can be boosted to higher
values.
Extreme collimated flows from collapsars with bulk 
Lorentz factors $>$ 100 have been proposed 
as sources of $\gamma$-ray bursts (e.g. Dar 1998; M\'esz\'aros \& Rees 1997; Fargion, 1998).
Because the jets are highly directional, the properties of the bursts 
will depend on the viewing angle relative to the rotation axis. In this context, 
the study of less extreme collimated flows in our own Galaxy may provide clues 
for a better understanding of the super-relativistic jets associated to the more 
distant $\gamma$-ray bursters. Of particular interest are the afterglow phenomena 
recently observed in microquasars that 
may result from internal shocks in the jets or from the impact of these jets in 
the environmental interstellar matter. However, $\gamma$-ray bursters are different to the 
microquasars found so far in our own Galaxy. The former do not repeat and 
seem to be related to catastrophic 
events, and have much larger super-Eddington luminosities. Therefore, 
the scaling laws in terms of the black hole mass that are valid in the analogy 
between microquasars and quasars do not apply in the case of $\gamma$-ray bursters. 

\vfill\eject

\vsl
\ni 12. CONCLUSIONS AND PERSPECTIVES
\vsl

The study of relativistic jets from X-ray binaries in our own galaxy 
sets on a firmer basis the relativistic ejections seen elsewhere 
in the Universe. The analogy between quasars and microquasars lead 
to the discovery of superluminal sources in our own galaxy, where it is 
possible to follow the motions of the two-sided ejecta. This permits  
to overcome the ambiguities that had dominated the physical interpretation 
of one-sided moving jets in quasars, and 
conclude that the ejecta consist 
mainly of matter moving with relativistic bulk motions, rather than waves 
propagating through a slowly moving jet. The Lorentz factors 
of the bulk motions in the jets from
microquasars seem to be similar to those believed to be common in quasars. 
From the study of the 
two-sided moving jets in one microquasar, an upper limit for 
the distance to the source was derived, using constraints from special relativity.

Because of the relative short timescales of the phenomena associated 
with the flows of matter around stellar mass black holes, one can 
sample phenomena that we have not been able to observe in quasars. Of particular 
importance is to understand the connection between accretion flow instabilities observed 
in the X-rays, with the ejection of relativistic clouds of plasma observed in 
the radio, infrared, and possibly in the optical. The detection of synchrotron 
infrared flares implies that the ejecta in microquasars contain very 
energetic particules with Lorentz factors of at least 10$^3$.

The discovery of microquasars opens several new perspectives that could prove 
to be particularly productive:

1. They provide a new method to determine distances using special relativity constraints. 
If the proper motions of the two-sided ejecta and the Doppler factor of 
a spectral line from one ejecta are measured, the distance to the source
can be derived.
With the rapid advance of technological capabilities in astronomy, this 
relativistic method to determine distances may be applied first to black hole 
jet sources in galactic binaries, and in the decades to come to quasars. 
 
2. Microquasars are nearby laboratories that can
be used to gain a general understanding of the mechanism of ejection 
of relativistic jets. The multiwavelength observations of GRS 1915+105 
during large-amplitude oscillations suggest that the clouds are 
ejected during the replenishment of the inner accretion disk that follows 
its sudden dissappearance beyond the last stable orbit around the black hole. 
In the context of these new data, the time seems to be ripe for new theoretical 
advances on the models of formation of relativistic jets. 

3. High sensitivity X-ray spectroscopy of jet sources with 
future X-ray space observatories may clarify the phenomena in accretion disks 
that are associated to the formation of jets. 

4. More microquasars will be discovered in the future. 
Among them, microblazars should appear as sources with fast and 
large amplitude variations in the observed flux. Depending on the beaming angle and bulk 
Lorentz factor they will be observed up to very high photon energies.

5. The spin of stellar mass black holes could be derived from the 
observed maximum stable frequency of the QPOs observed in the
X-rays, provided the mass has been independently determined.
However, theoretical 
work is 
needed to distinguish between the alternative interpretations 
that in the context of general relativity have been proposed for the
maximum stable frequency of QPOs.

6. Finally, microquasars
could be test grounds for general relativity theory in the strong field limit.
General relativity theory in weak gravitational fields has been successfully 
tested by observing in the radio wavelengths
the expected decay in the orbit of a 
binary pulsar, an effect produced by gravitational radiation
damping. We expect that phenomena observed in microquasars could be used 
in the future to investigate the physics
of strong field relativistic gravity near the horizon of black holes.

\vsl
\ni ACKNOWLEDGEMENTS
\vsl

We are most grateful to Vivek Dhawan for permission to include in this review 
unpublished results from our VLBA observations, and to Sylvain Chaty and Josep 
Mart\'\i\ for help 
in producing Figure 7. We thank Ralph Spencer, Jacques Paul, 
Josep Mart\'\i, and Alan Harmon for comments on the original manuscript. 
We are also greateful to Philippe Durouchoux for the organization 
of the workshop on {\it Relativistic Jet Sources in the Galaxy} held in 
Paris on December 12-13, 1998, from which 
we have benefitted. During this work, LFR received partial
support from CONACyT, M\'exico and DGAPA, UNAM.

\vfill\eject

\ni {\it Figure Captions}
\vsl

Figure 1. Contour map of the 6-cm emission from the radio counterpart of
1E1740.7-2942, as observed with the Very Large Array
(Mirabel et al. 1992; Rodr\'\i guez \& Mirabel 1999c). The error circle of the ROSAT
position (Heindl, Prince, \& Grunsfeld 1995), that includes the core source,
is also shown. At a distance of 8 kpc the length of the jet structure would be 
$\sim$5 pc. The half power contour of the beam is shown in the top left corner.
Contours are -4, 4, 5, 6, 8, 10, 12, 15, and 20 times 28 $\mu$Jy
beam$^{-1}$.

Figure 2. Pair of radio condensations moving away from the hard X-ray source GRS 1915+105 
(Mirabel and Rodr\'\i guez, 1994). 
These uniform-weight VLA maps were made at $\lambda$3.5-cm for the 1994 epochs on the 
right side of each map. 
The position of the stationary core is indicated with a small cross. 
The maps have been rotated 60$^{\circ}$ clockwise for easier display. 
The cloud to the left appears to move away from the stationary core
at 125\% the speed of light. Contours are 1, 2, 4, 8, 16, 32, 64, 128, 256 
and 512 times 0.2 mJy/beam for all epochs except for March 27 where the contour 
levels are in units of 0.6 mJy/beam. 
The half power beam width of the observations, 0.2 arc sec, is shown in the top 
right corner. 

Figure 3. Angular displacements as a function of time
for four ejection events observed in 1994 in GRS~1915+105
(Rodr\'\i guez \& Mirabel 1999a).
Top: Angular displacements as a function of time
for four approaching condensations corresponding to
ejections that took place on (from left to right)
1994 January 29 (triangles), February 19 (squares), March 19 (circles), 
and April 21 (crosses).
Bottom: Angular displacements as a function of time
for three receding condensations corresponding to
ejections that took place on (from left to right)
1994 February 19 (squares), March 19 (circles), and April 21 
(crosses). The clouds of the 1994
January 29 ejection were relatively weak and the receding
component could not be
detected unambiguously. The dashed
lines are the least squares fit to the angular displacements
of the 1994 March  19 event, the brighter and better
studied. Note that the motions appear to be ballistic (that is, unaccelerated).

Figure 4. Contour map of the 2-cm emission from
the core of GRS~1915+105, as observed on April 11, 1997 with the Very Long 
Baseline Array
at milliarcsecond angular resolution (Dhawan, Mirabel and Rodr\'\i guez 1999).
The angular resolution corresponds to about 10 AU at
GRS~1915+105.
The half power contour of the beam is shown in the bottom left corner.
Contours are -1, 1, 2, 4, 8, 16, 32, 64, and 96 times 0.26 mJy
beam$^{-1}$. The position angle of this ejection at milliarcsec scale,
is the same as that seen at the arcsec scales three years before.

Figure 5. A sequence of seven VLBA images of GRO~J1655--40 at 1.6 GHz, each
rotated anticlockwise by 43$^\circ$, and each having an angular resolution of 
3.0$\times$0.4 arcsec (Hjellming \& Rupen 1995). Each image is 
labeled with the date of the observations.
The solid lines between images identify motions of 54 mas day$^{-1}$ (left)
and 45.5 mas day$^{-1}$ (right). The vertical line marks the position of the central 
source, assumed to be the brightest point on each image.

Figure 6. Geometry of the two-sided ejection. 
The emission is symmetric, but when the emitting clouds move at
relativistic speeds the approaching component of the pair appears to move faster and
to be brighter than the receding component.

Figure 7. Radio, infrared, and X-ray light curves
for GRS~1915+105 at the time of quasi-periodic
oscillations on
1997 September 9 (Mirabel et al. 1998).
The infrared flare starts during the recovery from the X-ray dip,
when a sharp, isolated X-ray spike is observed.
These observations show the connection between the rapid disappearance
and follow-up replenishment of the inner accretion disk seen in the
X-rays (Belloni et al. 1997), and the ejection of relativistic plasma clouds 
observed as synchrotron 
emission at infrared wavelengths first and later at radio wavelengths.
A scheme of the relative positions where the different emissions originate
is shown in the top part of the figure.
The hardness ratio
(13-60 keV)/(2-13 keV) is shown at the bottom of the figure.

Figure 8. Very Large Array continuum mosaic of W50 at 1.5 GHz (Dubner et al. 1998). 
The radio counterpart of SS433 is the bright unresolved source at the center of 
the image. The lateral E-W extension of the nebula over $\sim$1$^{\circ}$ ($\sim$50 pc) 
is caused by the injection of the relativistic jets from SS433. 
The greyscale varies from 1 to 25 mJy beam$^{-1}$.
The angular resolution is 56$\times$54 arcsec.

\vfill\eject

\ni {\it Literature Cited}
\vsl

\frenchspacing

\nhi Ables, JG. 1969, {\it Proc. Astron. Soc. Australia} 1: 237-40

\nhi Abramowicz, MA, Chen, X, Kato, S, Lasota, JP, Reguev,
O. 1995, {\it Ap. J.} 438: L37-40

\nhi Atoyan, AM, Aharonian, FA. 1997,
{\it Ap. J.} 490: L149-52

\nhi Ball, L, Kesteven, MJ, Campbell-Wilson, D, Turtle, AJ, Hjellming, RM.
1995, {\it MNRAS} 273: 722-30 

\nhi Ball, L. 1996, in {\it Radio Emission from the Stars and
the Sun, ASP Conf. Ser.} 93: 219-27

\nhi Ballet, J. et al. 1994, {\it AIP Conference Proceedings} 308: 131-34

\nhi Barret, D, McClintock, JE, Grindlay, JE. 1996, {\it Ap. J.} 473: 963-73

\nhi Begelman, MC. 1998, {\it Ap. J.} 493: 291-300

\nhi Begelman, MC, Blandford, RD. 1998, in {\it Workshop on Relativistic Jet 
Sources in the Galaxy}. Paris, December 12-13, 1998

\nhi Begelman, MC, Hatchett, SP, McKee, CF, Sarazin, CL, Arons, J. 1980,
{\it Ap. J.} 238: 722-30 

\nhi Belloni, T, M\'endez, M, King, AR, van der Klis, M,
van Paradijs, J. 1997, {\it Ap. J.} 479: L145-48 

\nhi Blandford, RD, Ostriker, JP. 1978, {\it Ap. J.} 221: L29-32 

\nhi Blandford, RD, Payne, DG. 1982, {\it MNRAS} 199: 883-94

\nhi Blandford, RD, Znajek, RL. 1977, {\it MNRAS} 179: 433-40 

\nhi Bodo, G, Ghisellini, G. 1995, {\it Ap. J.} 441: L69-71

\nhi Bridle, AH, Perley, RA. 1984, {\it Annu. Rev. Astr. Astrophys.} 22:
319-58

\nhi Brinkmann, W, Aschenbach, B, Kawai, N. 1996,
{\it Astron. Astrophys.} 312: 306-16

%\nhi Carilli, CL, Bartel, N, Diamond, P. 1994,
%{\it Astron. J.} 108: 64 

\nhi Castro-Tirado, AJ, Brandt, S, Lund, N, Lapshov, I, Sunyaev, RA, et al.
1994, {\it Ap. J. Suppl.} 92: 469-72

\nhi Castro-Tirado, AJ, Geballe, TR, Lund, N. 1996,
{\it Ap. J.} 461: L99-102

\nhi Chaty, S, Mirabel, IF, Duc, PA, Wink, JE, Rodr\'\i guez, LF. 1996, 
{\it Astron. Astrophys.} 310: 825-30

\nhi Chaty, S, Mirabel, IF, Rodr\'\i guez, LF. 1999,
in preparation

\nhi Chen, X, Swank, JH, Taam, RE. 1997,
{\it Ap. J.} 477: L41-44

\nhi Churazov, E, Gilfanov, M, Sunyaev, R, Khavenson, N, Novikov, B,
et al. 1994, {\it Ap. J. Suppl.} 92: 381-85

\nhi Cordier, B, Paul, J, Ballet, J, Goldwurm, A, Bouchet, L, et al,
1993, {\it Astron. Astrophys.} 275, L1-4

\nhi Cowley, AP, Schmidthe, PC, Crampton, D, Hutchings, JB. 1998, {\it Ap. J.} 504: 854-65

\nhi Cui, W, Zhang, SN, Chen, W. 1998, {\it Ap. J.} 492: L53-56

\nhi Curtis, HD. 1918, {\it Publ. Lick Obs.} 13: 9-42

\nhi Dar, A. 1998, {\it Ap. J.} 500: L93-96

\nhi Dela Valle, M, Mirabel, IF, Rodr\'\i guez, LF. 1994,
{\it Astron. Astrophys.} 290: 803-06

\nhi Dhawan, V, Mirabel, IF, Rodr\'\i guez, LF. 1999, in preparation 

\nhi Djorgovski, S, Thompson, D, Mazzarella, J, Klemola, A, Neugebauer, G.
1992, {\it IAU Circular 5596}

\nhi Dubner, GM, Holdaway, M. Goss, WM, Mirabel, IF. 1998, {\it Astron. J.} 116: 1842-55

\nhi Ebisawa, K. 1996, in {\it X-ray Imaging and Spectroscopy of Cosmic Hot Plasmas}, ed. F. 
Makino \& K. Mitsuda, Tokyo: University Academy Press, 427-31

\nhi Eckart, A, Genzel, R. 1997, {\it MNRAS} 284: 576-98

\nhi Eikenberry, SS, Matthews, K, Morgan, EH, Remillard, RA, Nelson, RW.
1998a, {\it Ap. J.} 494: L61-64

\nhi Eikenberry, SS, Matthews, K, Murphy, TW, Nelson, RW, Morgan, EH, Remillard, RA, Muno, M.
1998b, {\it Ap. J.} 506: L31-34

\nhi Fabian, AC, Rees, MJ. 1979,
{\it MNRAS} 187: 13p-16p

\nhi Falcke, H, Mannheim, K, Biermann, PL. 1993,
{\it Astron. Astrophys.} 278: L1-4 

\nhi Fargion, D. 1998, astro-ph/9808005

\nhi Fejes, I. 1986, {\it Astron. Astrophys.} 168: 69-71

\nhi Fender, RP, Bell Burnell, SJ, Williams, PM, Webster, AS. 1996,
{\it MNRAS} 283: 798-804

\nhi Fender, RP, Garrington, ST, McKay, DJ, Muxlow, TWB, Pooley, GG,
 Spencer, RE, Stirling, AM, Waltman, EB. 1999, {\it MNRAS} in press 

\nhi Fender, RP, Pooley, GG. 1998, {\it MNRAS} 300: 573-76

\nhi Fender, RP, Pooley, GG, Brocksopp, C, Newell, SJ. 1997,
{\it MNRAS} 290: L65-69

\nhi Fender, RP, Spencer, R, Tzioumis, T, Wu, K. et al. 1998, {\it Ap. J.} 506: L121-125

\nhi Finogenov, A. et al. 1994, {\it Ap. J.} 424: 940-42

\nhi Fomalont, EB, Geldzahler, BJ. 1991,
{\it Ap. J.} 383: 289-94 

\nhi Foster, RS, Waltman, EB, Tavani, M, Harmon, BA, Zhang, SN, et al.
1996, {\it Ap. J.} 467: L81-84

\nhi Garc\'\i a, MR et al. 1998, in {\it Workshop on Relativistic Jet 
Sources in the Galaxy}. Paris, December 12-13, 1998
 
\nhi Ghosh, P, Abramowicz, MA. 1997,  {\it American Astron. Soc.} 191, 66

\nhi Gilmore, WS., Seaquist, ER.
1980, {\it Astron. J.} 85: 1486-95

\nhi Gilmore, WS., Seaquist, ER, Stocke, JT, Crane, PC.
1981, {\it Astron. J.} 86: 864-70

\nhi Greiner, J, Morgan, EH, Remillard, RA. 1996, {\it Ap. J.} 473: L107-10 

\nhi Grove, JE, Johnson, WN, Kroeger, RA, McNaron-Brown, K, Skibo, JG, et al.
1998, {\it Ap. J.} 500: 899-908

\nhi Hameury, JM, Lasota, JP, McClintock, JE, Narayan, R.
1997, {\it Ap. J.} 489: 234-43

\nhi Harmon, BA, Deal, KJ, Paciesas, WS, Zhang, SN, Gerard, E, 
Rodr\'\i guez, LF, Mirabel, IF. 1997, {\it Ap. J.} 477: L85-90

\nhi Harmon, BA, Zhang, SN, Wilson, CA, Rubin, BC, Fishman, GJ, et al.
1994, in {\it AIP Conference Proceedings No. 304} eds. Fichtel, CE, Gehrels, N,
Norris, JP. (AIP: New York), 210-19 

\nhi Heindl, WA, Prince, TA, Grunsfeld, JM. 1995, 
{\it Ap. J.} 430: 829-33

\nhi Hjellming, RM, Han, X. 1995, in {\it X-Ray Binaries 
(Cambridge University Press: Cambridge)}, p. 308

\nhi Hjellming, RM, Johnston, KJ. 1981,
{\it Ap. J.} 246: L141-45

\nhi Hjellming, RM, Johnston, KJ. 1988,
{\it Ap. J.} 328: 600-09

\nhi Hjellming, RM, Mioduszewski, AM. 1998, {\it IAU Circular 6872}

\nhi Hjellming, RM, Rupen, MP. 1995, {\it Nature} 375: 464-67

\nhi Hjellming, RM, Rupen, MP,  Mioduszewski, AM, et al. 1998, in 
{\it Workshop on Relativistic Jet 
Sources in the Galaxy}. Paris, December 12-13, 1998

\nhi Hunstead, RW, Wu, K, Campbell-Wilson, D. 1998, in preparation

\nhi Kalogera, V, Baym, G. 1996, {\it Ap. J.} 470: L61-64

% \nhi Kellermann, KI, Owen, FN. 1988, in {\it Galactic and 
% Extragalactic Radio Astronomy}, Springer, New York, eds 
% Verschuur, G.L. \& Kellermann, K.I.
\nhi King, A. 1998, in {\it Workshop on Relativistic Jet 
Sources in the Galaxy}. Paris, December 12-13, 1998

\nhi Koide, S., Shibata, K., Kudoh, T.
1998, {\it Ap. J.} 495: L63-66

\nhi Kudoh, T. \& Shibata, K. 1995,
{\it Ap. J.} 452: L41-44

\nhi Kuznetsov, S, Gilfanov, M, Churazov, E, Sunyaev, R, Korel, I, et al.
1997, {\it MNRAS} 292: 651-56 

\nhi Laurent, P, Paul, J. 1994, {\it Ap. J. Suppl.} 92: 375-79

\nhi Livio, M. 1998 in {\it Accretion Flows and Related Phenomena}
IAU Colloquium 163, eds. Wickramasinghe, D, Ferrario, L, Bicknell, G.
in press

\nhi Livio, M., Ogilvie, GI, Pringle, JE. 1998, {\it Ap. J.} in press

\nhi Lo, KY, Shen, ZQ, Zhao, JH, Ho, PTP. 1998,
{\it Ap. J.} 508, L61-64

\nhi L\'opez, JA. 1997,  {\it Planetary Nebulae},
IAU Symposium No. 180 pp. 197-203, eds. H.J. Habing 
\& H.J.G.L.M. Lamers, Kluwer.

\nhi Lucek, SG, Bell, AR. 1997, {\it MNRAS} 290, 327-33 

\nhi Mahadevan, R. 1998, {\it Nature} 394: 651-53 

\nhi Margon, BA, Stone, RPS, Klemola, A, Ford, HC, Katz, JI, et al.
1979, {\it Ap. J.} 230: L41-45

\nhi Margon, BA. 1984, {\it Annu. Rev. Astr. Astrophys.} 22: 507-36 

\nhi Mart\'\i, J, et al. 1999, in preparation

\nhi Mart\'\i, J, Mereghetti S, Chaty, S, Mirabel, IF, Goldoni, P, et al.
1998, {\it Astron. Astrophys.} 338: L95-99

\nhi Mart\'\i, J, Paredes, JM, Estalella, R. 1992, {\it Astron. Astrophys.} 258: 309-15

\nhi Mart\'\i, J, Rodr\'\i guez, LF, Mirabel, IF, Paredes, JM. 1996,
{\it Astron. Astrophys.} 306: 449-54

\nhi Matt, G. et al. 1998, in {\it Workshop on Relativistic Jet 
Sources in the Galaxy}. Paris, December 12-13, 1998 

\nhi Meier, DL, Edgington, S, Godon, P, Payne, DG, 
Lind, KR. 1997, {\it Nature} 388: 350-52

\nhi M\'esz\'aros, P, Rees, MJ. 1997,
{\it Ap. J.} 482: L29-32

\nhi Milgrom, M. 1979, {\it Astron. Astrophys.} 79: L3-6

\nhi Mioduszewski, AM. et al. 1998, in {\it Workshop on Relativistic Jet 
Sources in the Galaxy}. Paris, December 12-13, 1998 

\nhi Mirabel, IF, Bandyopadhyay, R, Charles, PA, Shahbaz, T, Rodr\'\i guez, LF.
1997, {\it Ap. J.} 477: L45-48

\nhi Mirabel, IF, Claret, A, Cesarsky, CJ, Cesarsky, DA, Boulade, O. 1996b,
{\it Astron. Astrophys.} 315: L113-16

\nhi Mirabel, IF, Dhawan, V, Chaty, S, Rodr\'\i guez, LF, 
Robinson, C, Swank, J, Geballe, T. 1998,
{\it Astron. Astrophys.} 330: L9-12

\nhi Mirabel, IF, Duc, P-A, Rodr\'\i guez, LF. et al. 1994, 
{\it Astron. Astrophys.} 282: L17-20

%\nhi Mirabel, IF, Rodr\'\i guez, LF. 1995, in {\it Proceedings of the 
%XVII Texas Symp. on Relativistic Astrophysics and Cosmology,} Annals 
%of the New York Academy of Sciences, 759, p. 21-37 

\nhi Mirabel, IF, Rodr\'\i guez, LF. 1994, {\it Nature} 371: 46-48

\nhi Mirabel, IF, Rodr\'\i guez, LF. 1998, {\it Nature} 392: 673-76 

\nhi Mirabel, IF, Rodr\'\i guez, LF, Chaty, S, Sauvage, M,
Gerard, E, et al. 1996a,
{\it Ap. J.} 472: L111-14

\nhi Mirabel, IF, Rodr\'\i guez, LF, Cordier, B, Paul, J, Lebrun, F.
1992, {\it Nature} 358: 215-17

\nhi Mirabel, IF, Rodr\'\i guez, LF, Cordier, B, Paul, J, Lebrun, F.
1993, in {\it Sub-arcsecond Radio Astronomy}, eds. RJ Davis \& RS Booth,
Cambridge University Press, 47-49

\nhi Morgan, EH, Remillard, RA, Greiner, J. 1997, {\it Ap. J.} 482: 993-1010 

\nhi Motch, C. 1998, {\it Astron. Astrophys.} 338: L13-16

\nhi Narayan, R, Garc\'\i a, MR, McClintock, JE. 1997, {\it Ap. J.} 478: L79-82

\nhi Narayan, R, Mahadevan, R, Grindlay, JE, Popham, RG, Gammie, C.
1998, {\it Ap. J.} 492: 554-68

\nhi Nowak, MA, Wagoner, RV, Begelman, MC, Lehr, DE. 1997, {\it Ap. J.} 477: L91-94

\nhi Orosz, JA, Bailyn, CD. 1997, {\it Ap. J.} 477: 876-96

\nhi Orosz, JA, Remillard, RA, Bailyn, CD, McClintock, JE. 1997,
{\it Ap. J.} 478: L83-86

\nhi Pacholczyk, AG. 1970, {\it Radio Astrophysics}, Freeman, San Francisco

\nhi Paragi, Z, Vermeulen, RC, Fejes, I, Schilizzi, RT, Spencer, RE, Stirling, AM. 
1998, in {\it Workshop on Relativistic Jet 
Sources in the Galaxy}. Paris, December 12-13, 1998  

\nhi Paul, J. et al. 1991, in {\it Advances in Space Research}, 11: 8289-302

\nhi Pearson, TJ, Zensus, JA. 1987, in {\it Superluminal Radio Sources,
Cambridge University Press}, eds.
Zensus, J. A. \& Pearson, T. J., p. 1  

\nhi Peebles, PJE. 1993, {\it Principles of Physical Cosmology}, Princeton
University Press, Princeton

\nhi Penrose, R. 1969, {\it Nuovo Cimento} 1, 252-76  

\nhi Phillips, SN, Shahbaz, T, Podsiadlowski, Ph. 1999, {\it MNRAS}, in press

\nhi Pooley, GG, Fender, RP. 1997, {\it MNRAS} 292: 925-33

\nhi Ramaty, R, Leventhal, M, Chan, KW, Lingenfelter, RE. 1992, {\it Ap. J.} 392: L63-67

\nhi Rees, MJ. 1966, {\it Nature} 211: 468-70

\nhi Rees, MJ. 1982, {\it The Galactic Center} AIP: New York, pp 166-76

\nhi Rees, MJ. 1984, {\it Annu. Rev. Astr. Astrophys.} 22, 471-506

\nhi Rees, MJ. 1998, in {\it Black Holes and Relativistic Stars}, ed. Wald, RM,
University of Chicago, 79-101

\nhi Reipurth, B., Bertout, C. 1997,
{\it Herbig-Haro Flows and the Birth of Stars,} IAU
Symposium No. 182 (Kluwer)

\nhi Remillard, RA, Morgan, EH, McClintock, JE, Bailyn, CD, Orosz, JA, et al.
1998, in {\it Proc. 18th Texas Symposium on Relativistic Astrophysics} eds.
Olinto, A, Frieman, J, Schramm, D. (Singapore: World Scientific), in press 

\nhi Rodr\'\i guez, L. F., Gerard, E. Mirabel, I. F.,
G\'omez, Y., Vel\'azquez, A., 1995, {\it Ap. J. Suppl.} 101: 173-79

\nhi Rodr\'\i guez, LF, Mirabel, IF. 1997, {\it Ap. J.} 474, L123-25

\nhi Rodr\'\i guez, LF, Mirabel, IF. 1999a, {\it Ap. J.} in press

\nhi Rodr\'\i guez, LF, Mirabel, IF. 1999b, {\it Astron. Astrophys.} 340: L47-50

\nhi Rodr\'\i guez, LF, Mirabel, IF. 1999c, in preparation 

\nhi Rodr\'\i guez, LF, Mirabel, IF, Mart\'\i, J. 1992 {\it Ap. J.} 401: L15-18

% \nhi Rybicki, G. B., Lightman, A. P. 1979, {\it Radiative Processes in
% Astrophysics}, Wiley, New York

\nhi Sams, BJ, Eckart, A, Sunyaev, R. 1996 {\it Nature} 382: 47-49

\nhi Scaltriti, F, Bodo, G, Ghisellini, G, Gliozzi, M, Trussoni, E.
1997, {\it Astron. Astrophys.} 327: L29-31

\nhi Schalinski, CJ, Johnston, KJ, Witzel, A, Spencer, RE, Fiedler, R, et al.
1995, {\it Ap. J.} 447: 752-59

\nhi Seaquist, ER. 1993, {\it Reports on Progress in Physics} 56: 1145-208

\nhi Smith, DA, Levine, A, Wood, A. 1998, {\it IAU Circular 6932}

\nhi Spencer, RE, 1979, {\it Nature} 282: 483-84

\nhi Spencer, RE, 1984, {\it MNRAS}, 209: 869-79

\nhi Spruit, H. C., Foglizzo, T., Stehle, R.
1997, {\it MNRAS} 288: 333-42

\nhi Stewart, RT, Caswell, JL, Haynes, RF, Nelson, GJ. 1993,
{\it MNRAS} 261: 593-98

\nhi Sunyaev, R, Churazov, E, Gilfanov, M, et al.
1991, {\it Ap. J.} 383: L49-53 

\nhi Tanaka, Y, Nandra, K, Fabian, AC, Inoue, H, Otani, C. et al.
1995, {\it Nature} 375: 659-61

\nhi Tanaka, Y, Shibazaki, N. 1996, {\it Annu. Rev. Astr. Astrophys.} 34: 607-44 

% \nhi Taylor, GB. 1996, {\it Ap. J.} 470: 394- 

\nhi Taylor, JH, Weisberg, JM. 1982, {\it Ap. J.} 253: 908-20 

\nhi Taylor, JH, Wolszczan, A, Damour, T, Weisberg, JM. 1992,
{\it Nature} 355: 132-36

\nhi Tingay, SJ, Jauncey, DL, Preston, RA, Reynolds, JE, Meier, DL, et al.
1995, {\it Nature} 374: 141-43

\nhi van der Laan, H. 1966, {\it Nature} 211: 1131-33

\nhi van Paradijs, J. 1995, in {\it X-Ray Binaries} (Cambridge University
Press: Cambridge), p. 536

\nhi Velusamy, T, Kundu, MR. 1974,
{\it Astron. Astrophys.} 32: 375-90

\nhi Vermeulen, RC, Schilizzi, RT, Spencer, RE, Romney, JD, Fejes, I.
1993, {\it Astron. Astrophys.} 270: 177-88

% \nhi Walker, R. C., Romney, J. D. \& Benson, J. M.
% 1994, {\it Ap. J.} 430: L45

\nhi Ueda, Y. et al. 1998, in {\it Workshop on Relativistic Jet 
Sources in the Galaxy}. Paris, December 12-13, 1998

\nhi Zensus, JA. 1997, {\it Annu. Rev. Astr. Astrophys.} 35: 607-36

\nhi Zhang, SN, Mirabel, IF, Harmon, BA, Kroeger, RA, Rodr\'\i guez, LF, et al.
1997, {\it Proceedings of the Fourth Compton Symposium} ed. CD Dermer, MS Strickman, JD Kurfess,
(AIP: New York), 141-62

\nhi Zhang, SN, Wilson, CA, Harmon, BA, Fishman, GJ, \& Wilson, RB.
1994, {\it IAU Circular 6046}

\vfill\eject

% vsize should be 11.0 truecm
%\special{landscape}
\hsize=19truecm  \hoffset=-0.5truecm  \vsize=15.5truecm  \voffset=-1.5truecm
%\hsize=25truecm   \vsize=18truecm  \voffset=0.0truecm \hoffset -2.5cm

\def\dblbaselines{\baselineskip=12pt    \lineskip=0pt   \lineskiplimit=0pt}
\def\vsl{\vskip\baselineskip}   \def\vs{\vskip 6pt}

\nopagenumbers
\def\t#1{#1} 
\def\t#1{\empty}
    
\def\skipit{\hskip 7pt}  \def\ba{\kern -1pt}
\parskip = 0pt % default is \parskip = 0pt+1pt
\def\ts{\thinspace} \def\cl{\centerline}
\def\ni{\noindent}  \def\nhi{\noindent \hangindent=1.0truecm}
\def\nhhi{\noindent \hangindent=3.30truecm}  
\def\makeheadline{\vbox to 0pt{\vskip-30pt\line{\vbox to8.5pt{}\the
                               \headline}\vss}\nointerlineskip}

\def\footnoterule{\kern-3pt \hrule width \hsize \kern 2.6pt \vskip 3pt}
\output={\plainoutput}    \pretolerance=10000   \tolerance=10000
\def\sup1{$^{\rm 1}$} \def\sup2{$^{\rm 2}$}
\def\r0{$\rho_0$}   
\def\00{$\phantom{000000}$} \def\0{\phantom{0}} \def\bb{\kern -2pt}

\def\1{\phantom{1}}

\newdimen\sa  \def\sd{\sa=.1em \ifmmode $\rlap{.}$''$\kern -\sa$
                               \else \rlap{.}$''$\kern -\sa\fi}
\def\ss{\ifmmode ^{\prime\prime}$\kern-\sa$ \else $^{\prime\prime}$\kern-\sa\fi}
\def\mm{\ifmmode ^{\prime}$\kern-\sa$ \else $^{\prime}$\kern-\sa \fi}

\def\m31{M{\ts}31} \def\mm32{M{\ts}32} \def\mmm33{M{\ts}33}

\dblbaselines

% \pageno=4 \toppageno

%\input cittable.tex
%
%  VERSION DOCTORED BY JK FOR TABLES ONLY
%
%----------%
%  TABLES  %
%----------%
\def\endtable{\endgroup}
\def\tableheight{\vrule width 0pt height 8.5pt depth 3.5pt}
{\catcode`|=\active \catcode`&=\active 
    \gdef\tabledelim{\catcode`|=\active \let|=\vbar
                     \catcode`&=\active \let&=\nobar} }
\def\table{\begingroup
    \def\twidth{\hsize}
    \def\tablewidth##1{\def\twidth{##1}}
    \def\defaultheight{\vrule width 0pt height 8.5pt depth 3.5pt}
    \def\heightdepth##1{\dimen0=##1
        \ifdim\dimen0>5pt 
            \divide\dimen0 by 2 \advance\dimen0 by 2.5pt
            \dimen1=\dimen0 \advance\dimen1 by -5pt
            \vrule width 0pt height \the\dimen0  depth \the\dimen1
        \else  \divide\dimen0 by 2
            \vrule width 0pt height \the\dimen0  depth \the\dimen0 \fi}
    \def\spacing##1{\def\defaultheight{\heightdepth{##1}}}
    \def\nextheight##1{\noalign{\gdef\tableheight{\heightdepth{##1}}}}
    \def\end{\cr\noalign{\gdef\tableheight{\defaultheight}}}
    \def\zerowidth##1{\omit\hidewidth ##1 \hidewidth}    
    \def\hline{\noalign{\hrule}}
    \def\skip##1{\noalign{\vskip##1}}
    \def\bskip##1{\noalign{\hbox to \twidth{\vrule height##1 depth 0pt \hfil
        \vrule height##1 depth 0pt}}}
    \def\header##1{\noalign{\hbox to \twidth{\hfil ##1 \unskip\hfil}}}
    \def\bheader##1{\noalign{\hbox to \twidth{\vrule\hfil ##1 
        \unskip\hfil\vrule}}}
    \def\spanloop{\span\omit \advance\mscount by -1}
    \def\extend##1##2{\omit
        \mscount=##1 \multiply\mscount by 2 \advance\mscount by -1
        \loop\ifnum\mscount>1 \spanloop\repeat \ \hfil ##2 \unskip\hfil}
    \def\vbar{&\vrule&}
    \def\nobar{&&}
    \def\hdash##1{ \noalign{ \relax \gdef\tableheight{\heightdepth{0pt}}
        \toks0={} \count0=1 \count1=0 \putout##1\end 
        \toks0=\expandafter{\the\toks0 &\end} \xdef\piggy{\the\toks0} }
        \piggy}
    \let\e=\expandafter
    \def\putspace{\ifnum\count0>1 \advance\count0 by -1
        \toks0=\e\e\e{\the\e\toks0\e&\e\multispan\e{\the\count0}\hfill} 
        \fi \count0=0 }
    \def\putrule{\ifnum\count1>0 \advance\count1 by 1
        \toks0=\e\e\e{\the\e\toks0\e&\e\multispan\e{\the\count1}\leaders\hrule\
hfill}
        \fi \count1=0 }
    \def\putout##1{\ifx##1\end \putspace \putrule \let\next=\relax 
        \else \let\next=\putout
            \ifx##1- \advance\count1 by 2 \putspace
            \else    \advance\count0 by 2 \putrule \fi \fi \next}   }
\def\tablespec#1{
    \def\vdimens{\noexpand\tableheight}
    \def\tabby{\tabskip=0pt plus100pt minus100pt}
    \def\r{&################\tabby&\hfil################\unskip}
    \def\c{&################\tabby&\hfil################\unskip\hfil}
    \def\l{&################\tabby&################\unskip\hfil}
    \edef\templ{\noexpand\vdimens ########\unskip  #1 
         \unskip&########\tabskip=0pt&########\cr}
    \tabledelim
    \edef\body##1{ \vbox{
        \tabskip=0pt \offinterlineskip
        \halign to \twidth {\templ ##1}}} }

\cl{\null} \vfill

$$
\table
\tablewidth{18truecm}
\tablespec{\l\l\l\l\l\l}
\body{
\header{{\bf Table 1} \quad Sources of Relativistic Jets in the Galaxy$^{(1)}$}
\skip{10pt}
\hline
\skip{.2truecm}\hline
\skip{5pt}
& Source & Compact object & $V_{app}$$^{(2)}$ & $V_{int}$$^{(3)}$ & $\Theta^{(4)}$ & References & \end
\skip{5pt}
\hline
\skip{5pt}
& GRS 1915+105 & black hole & 1.2c-1.7c & 0.92c-0.98c & 66$^{\circ}$-70$^{\circ}$ & MR94; F+99; DMR99 & \end
\skip{2pt}
& GRO J1655-40 & black hole & 1.1c & 0.92c & 72$^{\circ}$-85$^{\circ}$ & T+95; HR95; OB97 & \end
\skip{2pt}
& XTE J1748-288 & black hole & 0.9c-1.5c & $>$0.9c  &  & H+98 & \end
\skip{2pt}
& SS 433 & neutron star ? & 0.26c & 0.26c & 79$^{\circ}$ & M84; S84 & \end
\skip{2pt}
& Cygnus X-3 & neutron star ? & $\sim$0.3c & $\sim$0.3c & $>$70$^{\circ}$ & 
S+93;  M+99 & \end
\skip{2pt}
& CI Cam & neutron star ? & $\sim$0.15c & $\sim$0.15c  & $>$70$^{\circ}$ & M+98; G+98 & \end
\skip{2pt}
& Circinus X-1 & neutron star & $\geq$0.1c & $\geq$0.1c & $>$70$^{\circ}$ & S+93; F+98 & \end
\skip{2pt}
& 1E1740.7-2942 & black hole &  & & & M+92; RM99c & \end
\skip{2pt}
& GRS 1758-258 & black hole &  & & & R+94 & \end
\skip{2pt}
& Sgr A$^*$ & black hole &  &   &   &  L+98 & \end
\skip{2pt}
\skip{2pt}
\hline
}
\endtable
$$

\vskip -10pt

$^{(1)}$Sources reported as of December 1998. 
\vskip .05in 
$^{(2)}$V$_{app}$ is the apparent speed of the highest velocity component of the ejecta.
\vskip .05in 
$^{(3)}$V$_{int}$ is the intrinsic velocity of the ejecta.
\vskip .05in 
$^{(4)}$$\Theta$ is the angle between the direction of motion 
of the ejecta with the line of sight.

\vfill

\cl{\null} \eject

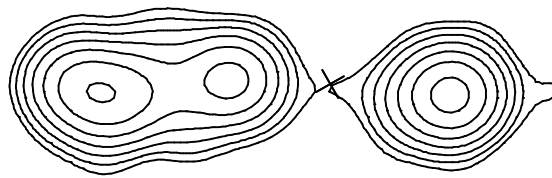
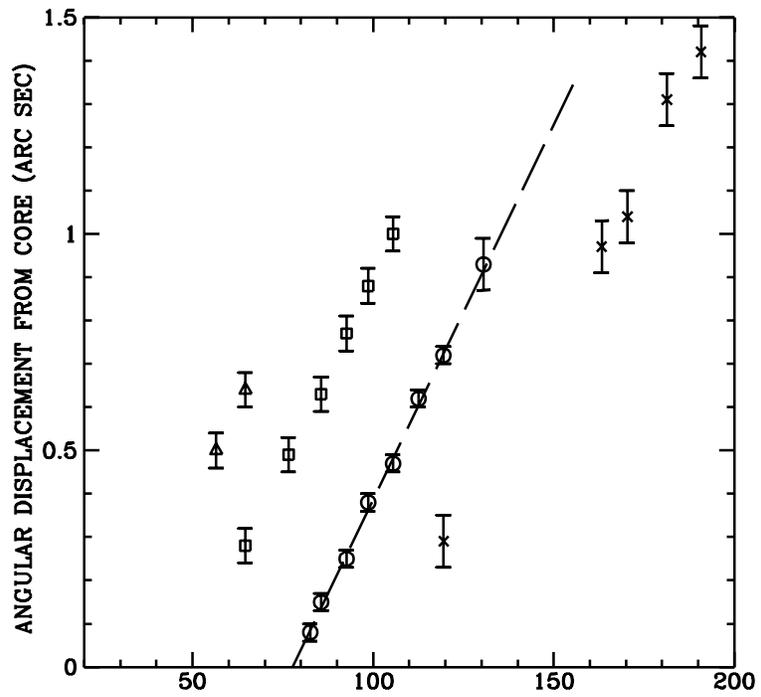
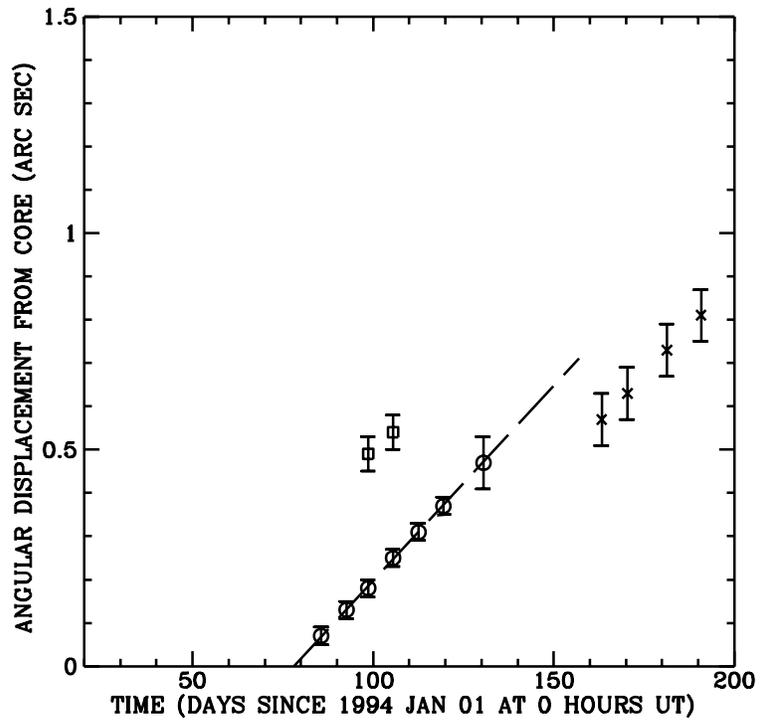
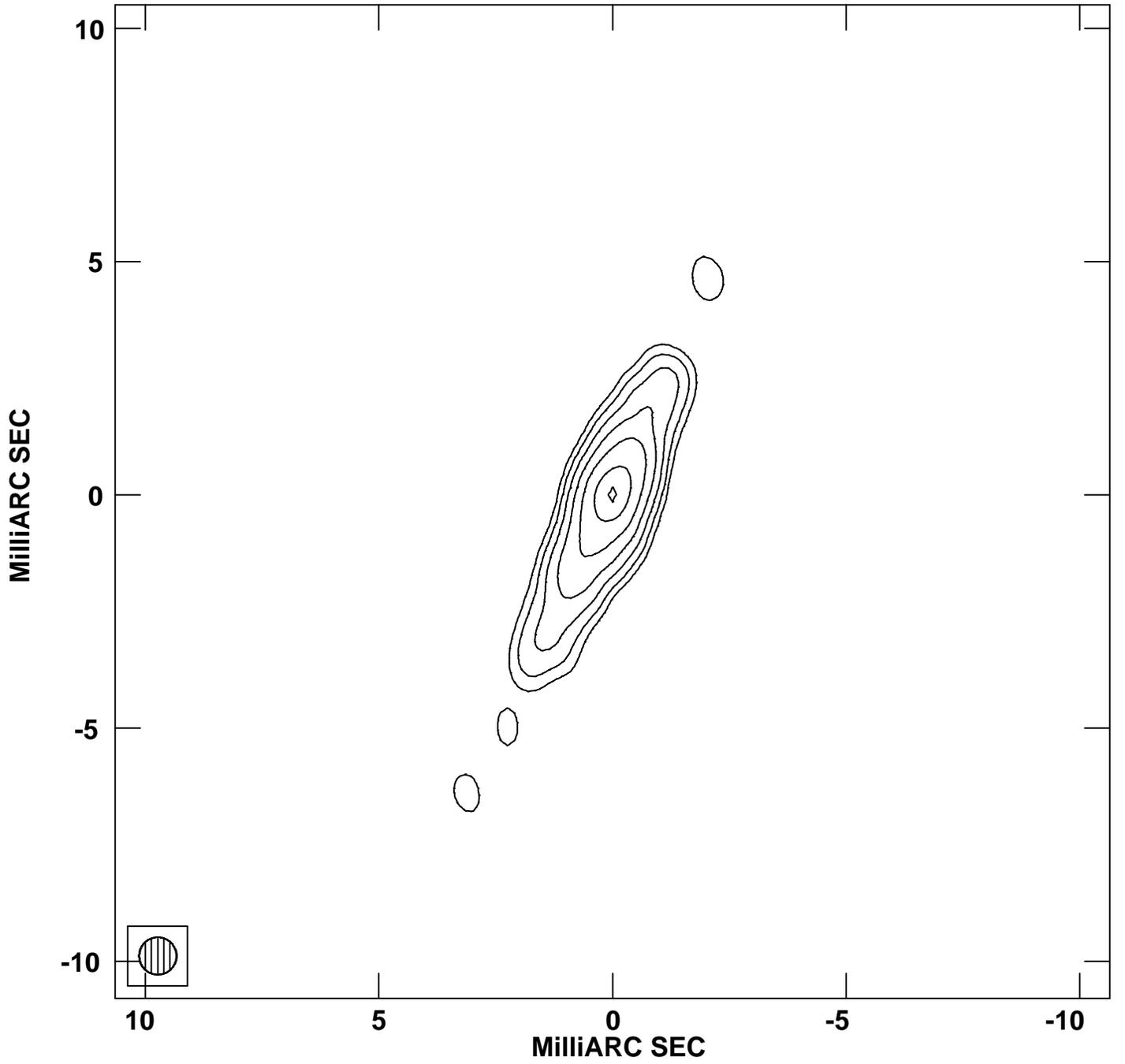
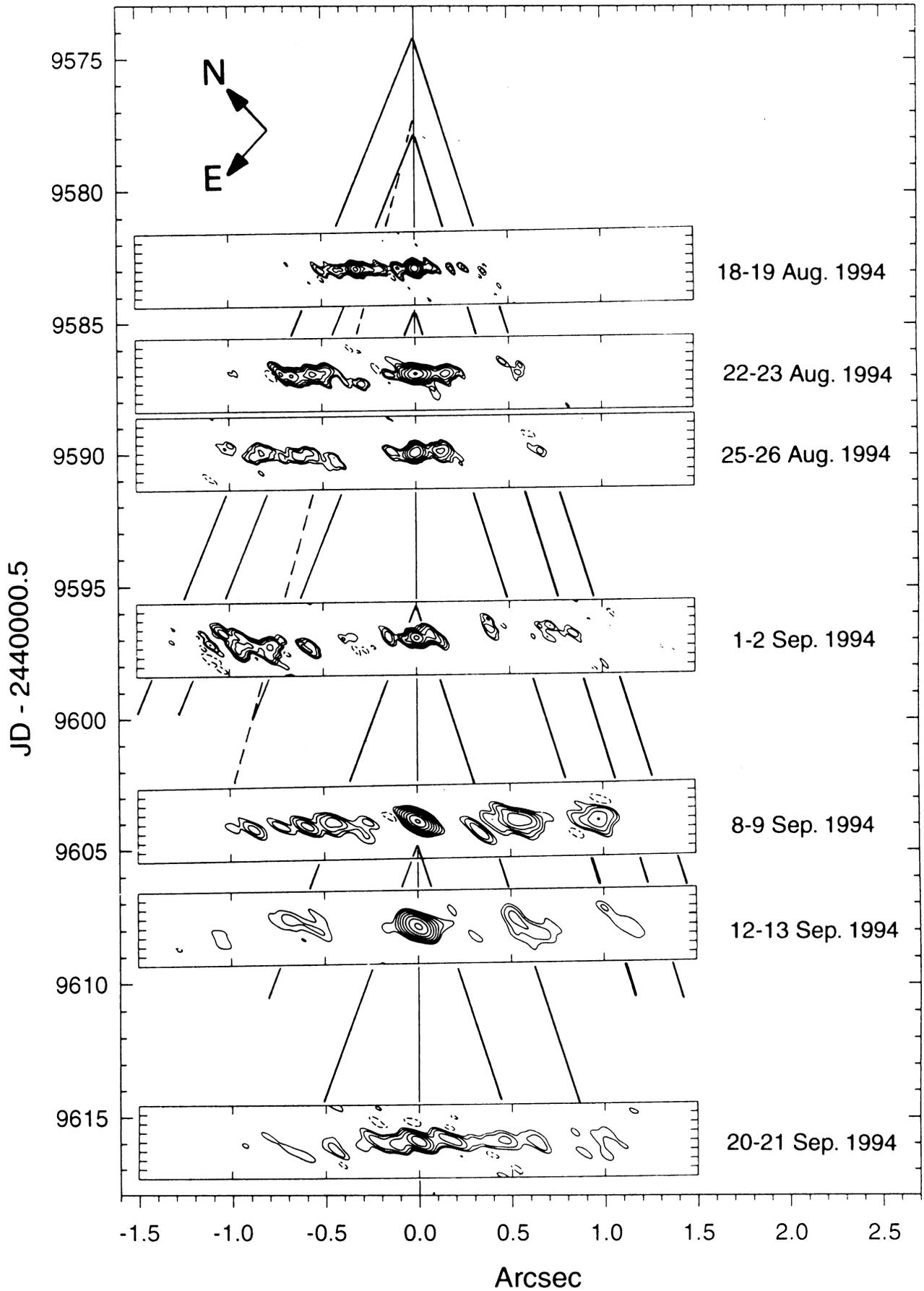
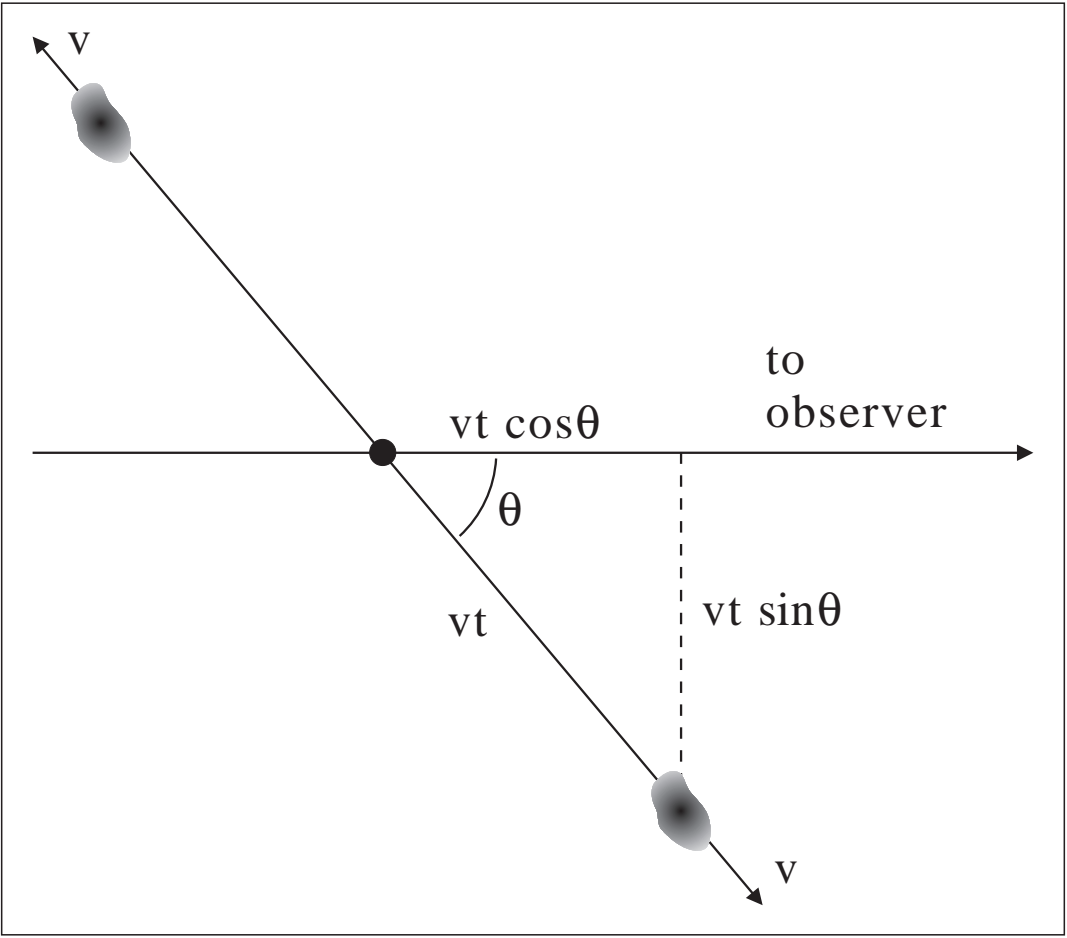
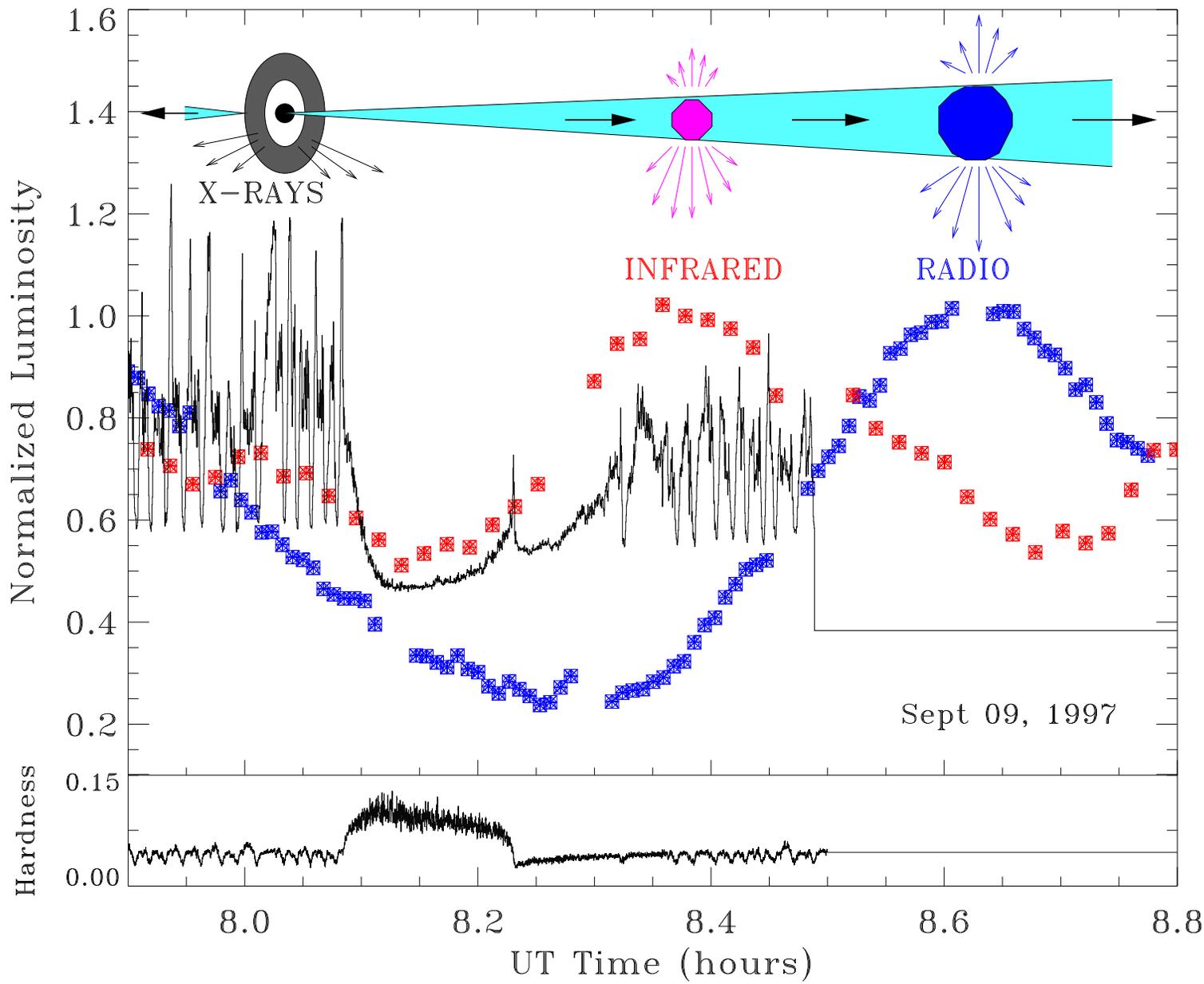
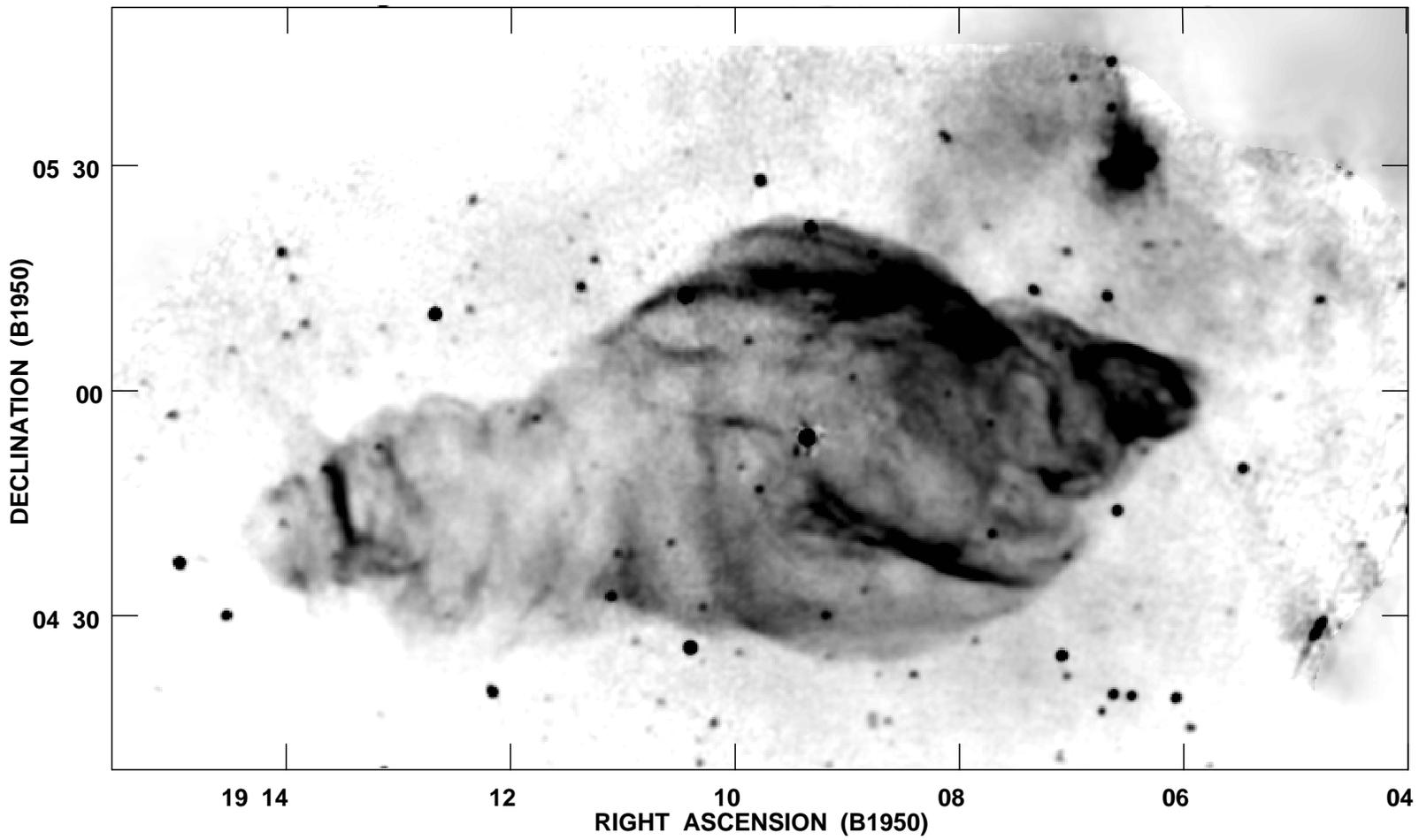
\end